\documentclass[hyper,a4paper]{JHEP3}

\usepackage{latexsym}
\usepackage{graphicx}
\usepackage{amssymb,amsmath}

\def \nqs#1#2{(\ref{#1})--(\ref{#2})}
\def \fig#1{Fig.~\ref{#1}}
\def \figs#1#2{Figs.~\ref{#1}--\ref{#2}}

\newcommand\bsm{\bar{B}_s\to\mu^+\mu^-}

\newcommand\bsg{\bar{B}\to X_s\gamma}
\newcommand\bbb{\bar{B}_s-B_s}

\newcommand\bsll{\bar{B}\to X_s l^+ l^-}

\newcommand\dll{\delta_{LL}^{d}} \newcommand\drr{\delta_{RR}^{d}}
\newcommand\dlr{\delta_{LR}^{d}} \newcommand\drl{\delta_{RL}^{d}}
\newcommand\ull{\delta_{LL}^{u}}

\newcommand\dxy{\delta_{XY}^{d}}

\newcommand\mdbare{m_{d}^{(0)}} 

\newcommand\mdphys{m_{d}} 

\newcommand\GdL{\Gamma_{d\,L}}
\newcommand\GdR{\Gamma_{d\,R}}

\newcommand\mgl{m_{\widetilde{g}}}
\newcommand\msq{m_{\widetilde{q}}}

\newcommand\brbsm{\mathrm{BR}(\bar{B}_s\to\mu^+\mu^-)}

\newcommand\brbsg{\mathrm{BR}(\bar{B}\to X_s\gamma)}
\newcommand\delmbs{\Delta M_{B_s}}

\newcommand\tanb{\tan\beta}

\newcommand\sutwo{\mathrm{SU}(2)}

\newcommand\ps{\,\mbox{ps}}
\newcommand\tev{\,\mbox{TeV}}
\newcommand\gev{\,\mbox{GeV}}

\newcommand\ie{\textit{i.e.}}

\title{Current and Future Limits on General Flavour Violation in
  {\boldmath $b\rightarrow s$} Transitions in Minimal Supersymmetry}

\author{John Foster${}^{a}$, Ken-ichi Okumura${}^b$ and Leszek Roszkowski${}^{c}$\\
${}^a$Dipartimento di Fisica, Universit\`a di Padova\\
Via F. Marzolo 8, I--35131, Padua, Italy\\
${}^b$Department of Physics, Kyushu University\\
Fukuoka 812--8581, Japan\\
${}^c$Department of Physics and Astronomy, University of Sheffield\\ 
Sheffield S3 7RH, UK
\email{john.foster@pd.infn.it, okumura@higgs.phys.kyushu-u.ac.jp, l.roszkowski@sheffield.ac.uk}}

\abstract{We discuss the current and prospective
limits that can be placed on supersymmetric contributions to $b\to s$
transitions in the Minimal Supersymmetric Standard Model with general
flavour mixing amongst the squarks. We consider three processes:
$\bsg$, $\bsm$ and $\bbb$ mixing, and pay particular attention to the
large $\tanb$ regime and beyond leading order contributions where the
difference between our analysis and previous analyses is most
pronounced. We find that even present limits on $\brbsm$ and $\delmbs$
often provide additional constraints on the amount of flavour
violation still allowed by $\brbsg$. Limits on supersymmetric and
Higgs mass parameters can also be strongly dependent on the flavour
violation. In particular, the case $\mu<0$ can still be allowed,
even for light sparticle masses. We
discuss how future measurements at the Tevatron, the LHC and elsewhere
can improve present limits but also 
provide unique signatures for the existence of flavour
violation.}

\preprint{{\sf KYUSHU--HET--87}}

\begin{document}

\section{Introduction}
\label{CUR:int}

The Minimal Supersymmetric Standard Model (MSSM) has, for many reasons,
become one of the most keenly investigated frameworks of particle physics beyond
the Standard Model (SM). Whilst its ultimate test
will be performed at the Large Hadron Collider (LHC), it is useful, in the 
meantime, to place limits on its parameters using existing
measurements. Examples of such measurements are the $W$ boson mass
and $\sin^2\theta_{\mathrm{eff}}$, the anomalous magnetic moment of the muon
and flavour changing neutral current (FCNC) processes. 

FCNC processes, in particular, provide a powerful constraint on the
MSSM, as the SM and supersymmetric (SUSY) contributions each enter at the
one--loop level. As the SM and SUSY contributions can be comparable
to one another, FCNC provide a unique means of probing the possible
flavour structure of the MSSM. Coupled with the increasingly stringent
limits that are emerging from the B factories and the Tevatron it
is natural to investigate how these constraints might affect the available
parameter space in the general MSSM and how such measurements might
complement direct searches for SUSY. In the most general formulation
of the squark soft terms -- namely general flavour mixing
(GFM) -- such measurements can become especially pertinent when
constraining the exact form the SUSY soft terms can take.

The usefulness of these constraints, however, is highly dependent on the
accuracy of the underlying calculation. It is therefore essential to
at least to attempt to calculate the SUSY contributions to FCNC 
processes at a similar level of accuracy as the existing SM
calculations~\cite{BRev}. In general, the additional particle
content of the MSSM makes such calculations rather prohibitive as
one must consider diagrams that, for example, feature gluinos as well
as gluons, and take care of possible mass hierarchies that exist in the
particle spectrum, resumming any large logarithms that might occur.
Complete next--to--leading order (NLO) calculations have therefore only
been completed in a few limiting cases~\cite{CDGG:bsg,BBKU:bsm} and
tend to focus on the more constrained assumption of minimal flavour
violation (MFV)~\cite{AGIS:bdec,Betal:MFV} rather than the GFM framework.
With these comments in mind it is useful to devise a formalism that builds
on existing leading order (LO) calculations to include the effects that
one might consider to be dominant beyond the leading order (BLO). Such
effects are typically considered to be proportional to either large
logarithms or $\tanb$ (the ratio of the vacuum expectation values of
the two Higgs doublets that appear in the MSSM). Large logarithms are
induced by hierarchies that might exist between the coloured SUSY
particles (the squarks and gluinos) and the characteristic electroweak
scale. The corrections proportional to $\tanb$, on the other hand, are
induced by threshold corrections to the down quark masses (the bottom
quark mass in particular)~\cite{Masses:th}, and the
charged~\cite{CH:th,DGG:bsg,CGNW:bsg} and neutral Higgs
vertices~\cite{NH:th,NH:th2,NH:rev}. In the GFM scenario, and without
adopting the mass insertion approximation (MIA), it has been
shown in~\cite{OR1:bsg,OR2:bsg,FOR:bdec} that the effects 
induced by the inclusion of these $\tanb$ enhanced threshold corrections
can be large, especially when compared to similar calculations
in the MFV scenario. Broadly speaking,
in the phenomenologically more viable region
$\mu>0$, the inclusion of such BLO corrections leads to a {\em focusing
effect} that can significantly loosen the bounds on SUSY sources of
flavour violation~\cite{OR1:bsg,OR2:bsg,FOR:bdec}. For $\mu<0$ the
situation is rather more complicated. On the one hand, the inclusion of
large logarithms can act to decrease certain supersymmetric contributions
to a given decay (in the case of $\bsg$ this is particularly true),
on the other, the $\tanb$ enhanced threshold corrections 
that acted to suppress the supersymmetric contributions for $\mu>0$
now act to increase them. Broadly speaking, however, it is possible,
once one includes GFM effects, to allow for
$\mu<0$, even for light sparticle masses (a region of parameter space
excluded in the MFV scenario), by considering relatively 
small sources of SUSY flavour violation~\cite{OR1:bsg,OR2:bsg}.

The aim of the paper is to present a discussion of the present
limits on SUSY flavour violation and the future
prospects that might arise from measurements at the Tevatron
and the LHC. We shall pay particular attention to the large
$\tanb$ regime where the differences between our analysis and existing
LO analyses in the literature~\cite{LO:bdec} are most pronounced. The
processes we shall use to constrain the various sources of flavour 
violation in the MSSM are: the inclusive decay $\bsg$; the rare 
decay $\bsm$; and the $\bbb$ mixing system. The SUSY contributions
to all three of these processes in the large $\tanb$ regime are known
to be sizeable.

The current world average for the branching ratio $\bsg$ is~\cite{HFAG}
\begin{align}
\label{brbsg:expt}
\brbsg_{\mathrm{exp}}=\left(3.39^{+0.30}_{-0.27}\right)\times 10^{-4}.
\end{align}
When compared with the current SM prediction for the branching ratio~\cite{NLO:bsg}
\begin{align}
\label{brbsg:sm}
\brbsg_{\mathrm{SM}}=\left(3.70\pm 0.30\right)\times 10^{-4},
\end{align}
it is clear that the contributions that arise from SUSY partner
exchange in both the MFV and GFM scenarios must be small
or suppressed in some way. A natural way this can be accomplished
in the GFM scenario is via the inclusion of BLO effects in the large
$\tanb$ regime. As discussed in~\cite{OR1:bsg,OR2:bsg,FOR:bdec},
when $\tanb$ is large and $\mu>0$, sizeable cancellations between
the chargino and gluino contributions to the decay can occur
once one has included the various $\tanb$ enhanced BLO corrections.
These cancellations inevitably lead to a suppression of the
overall supersymmetric contribution to the decay and can, in turn,
significantly loosen the bounds imposed by the decay in the large
$\tanb$ scenario~\cite{OR1:bsg,OR2:bsg,FOR:bdec}.

The other two processes we shall consider are currently unobserved.
The first is the rare decay $\bsm$ where only a upper bound exists
for the branching ratio. The current published $95\%$ confidence
limits are~\cite{CDF:bsm,D0:bsm}
\begin{align}
\label{bsmm:cdf:publ}
\brbsm_{\mathrm{CDF}}&<7.5\times 10^{-7},
\\
\label{bsmm:d0:publ}
\brbsm_{\mathrm{D{\O}}}&<5.0\times 10^{-7}.
\end{align}
Over the past year both CDF and D{\O} have provided preliminary results
for the limit on the branching ratio~\cite{CDF:bsmpr,D0:bsmpr}
\begin{align}
\label{bsmm:cdf:prel}
\brbsm_{\mathrm{CDF}}&<2.0\times 10^{-7},
\\
\label{bsmm:d0:prel}
\brbsm_{\mathrm{D{\O}}}&<3.8\times 10^{-7},
\end{align}
that may be combined to provide the improved
limit $\brbsm_{\mathrm{exp}}<1.5\times 10^{-7}$~\cite{Tev:bsm}.
Bearing in mind that the SM predicts~\cite{bsm:NLO}
\begin{align}
\brbsm_{\mathrm{SM}}=\left(3.46\pm 1.5\right)\times 10^{-9},
\end{align}
it is clear that the observation of the decay at the Tevatron would
provide a clear signal for physics beyond the SM. 

The second process we shall consider concerns the mixing between
the neutral $B^0_s$ mesons. The relevant observable here is the
mass difference between the mass eigenstates formed between the two mesons
$\delmbs$. Currently, only a lower bound exists for the quantity~\cite{HFAG}
\begin{align}
\delmbs^{\mathrm{exp}}>14.5\ps^{-1},
\end{align}
while the SM prediction is~\cite{bbb:NLO}
\begin{align}
\delmbs^{\mathrm{SM}}=\left(18.0\pm 3.7\right)\ps^{-1}.
\label{bbb:SM}
\end{align}
Comparing the current experimental limit with the SM prediction
it is clear that, should $\delmbs$ be measured just above the
current experimental limit, such values can be 
typically reconciled with the SM prediction~\eqref{bbb:SM}.
The more phenomenologically interesting region -- from the
point of view of observing new physics effects -- is, therefore,
where new physics induces contributions to $\delmbs$ far in
excess of the SM prediction. For example, if $\delmbs>30\ps^{-1}$,
such a measurement would imply a deviation from the SM at the
level of $2$--$3\sigma$~\cite{CKM}.

While only limits currently exist for both of these processes,
they already provide a useful limit in the large $\tanb$
regime due to the large enhancement that both processes receive. In
GFM, in particular, it has been shown that useful constraints
can be placed on SUSY flavour violation~\cite{FOR:bdec}
using these processes.

To summarise what follows, we first describe the bounds imposed by each
process in turn, discussing the currently allowed regions of
parameter space and how future measurements might affect this picture.
In sections~\ref{CUR:LimSing} and~\ref{CUR:Multi} we combine all three
constraints to find the allowed regions of parameter space for 
single and multiple sources of flavour violation respectively.

\section{Procedure}
\label{CUR:proc}

Before moving onto our numerical analysis it will be useful to briefly
overview the formalism that we will use to implement the BLO corrections
to the various decays considered in this paper, for more details we refer
the reader to~\cite{OR2:bsg,FOR:bdec} where the necessary ingredients
for such calculations are covered in great detail.

As discussed earlier in this paper, calculations of large effects that
appear beyond the leading order typically entails the inclusion of the
effects enhanced by either large logs of the form $\log M_{SUSY}/\mu_W$,
where $M_{SUSY}$ is a mass scale associated with the coloured superpartners
and $\mu_W$ is to be associated with the electroweak scale, or by $\tanb$.
One way of including both of these contributions is to consider an effective
theory where all the SUSY particles have been integrated out.
The  resulting theory is, essentially, a two Higgs doublet model that
includes the threshold 
corrections induced by integrating out the SUSY degrees of freedom. To do
this it is necessary to assume the following mass hierarchy
\begin{align}
M_{SUSY}\sim(\msq,\,\mgl)\gg\mu_W\sim(m_{H},\,m_W,\,m_t)\gg\mu_b\sim m_b.
\label{FORM:hier}
\end{align}
That is the squarks and gluino are heavier than the Higgs sector,
electroweak gauge bosons and top quark, which are in turn heavier
than the mass scale of the decays under consideration.
The remaining SUSY particles, such as the neutralinos and charginos,
are usually assumed to arise in the interval between $M_{SUSY}$ and
$\mu_W$.

When working in this effective field theory formalism $\tanb$ enhanced
corrections appear as threshold corrections to the masses
and couplings that appear in the theory defined at $\mu_W$. As an
example, consider the down quark mass matrix. In the effective
theory the down quark mass matrix $m_d$, which should be identified
with the physical mass matrix for the down quarks, is related
to the bare mass matrix $\mdbare$, defined in the full theory before
threshold corrections are taken into account, by the relation
\begin{align}
\mdphys=&\mdbare+\delta m_d={\mathrm{diag}}\left(m_d,m_s,m_b\right),
\end{align}
where $\delta m_d$ denotes the threshold corrections calculated
in the basis where $\mdphys$ is diagonal (the physical
super--CKM basis), exact expressions may be found in~\cite{FOR:bdec}.
The flavour structure of $\delta m_d$, and therefore $\mdbare$,
is highly non--trivial and, even in MFV, the matrices have off--diagonal
elements that are enhanced by $\tanb$, the inclusion of GFM effects
only serves to enhance these corrections. The non--trivial flavour
structure of the bare mass matrix acts as an additional source of
flavour violation through its appearance in the down squark mass matrix
and supersymmetric vertices, which are not subject to the threshold
corrections that affect the down quark mass matrix. Similar corrections
arise for the couplings of the quarks to the gauge and Higgs bosons,
the corrections to the Higgs boson couplings in particular, can
be appreciable even if $M_{SUSY}$ is large. The additional contributions
to flavour violating processes induced by these corrected couplings
and masses represent the $\tanb$ enhanced corrections that appear
beyond the leading order. It should be noted that when one works in
the mass eigenstate formalism, where one diagonalises the squark mass
matrix and flavour violation is mediated by the matrices $\GdL$ and
$\GdR$, it is necessary to employ an iterative procedure to accurately
calculate $\mdbare$ and the other vertices in the theory, the exact
details of this procedure are discussed in~\cite{OR2:bsg,FOR:bdec}.

Now that we have discussed the formalism we shall use let us now
discuss the remaining aspects of our calculation. After one
has calculated the bare mass matrix and corrected electroweak
vertices the supersymmetric contributions to the process under
consideration are evaluated, taking into account the effects
of the bare mass matrix, and evolved from the scale $M_{SUSY}$ 
to the electroweak scale $\mu_W$ using the relevant six flavour anomalous
dimension matrix. The remaining electroweak contributions (that
is the gauge boson and Higgs contributions) are then evaluated
using the uncorrected vertices when evaluating the NLO corrections
and the corrected vertices when evaluating the LO contributions.
The combined SUSY and electroweak contributions are finally evolved
to the scale $\mu_b$ and used to calculate the relevant observable
for the process in question.

Flavour violation in the soft breaking Lagrangian is often
parameterised in terms of the dimensionless
quantities
\begin{align}
\left(\dll\right)_{ij}=
&\frac{\left(m_{d,LL}^2\right)_{ij}}
{\sqrt{\left(m_{d,LL}^{2}\right)_{ii}\left(m^{2}_{d,LL}\right)_{jj}}},
&\left(\dlr\right)_{ij}=
&\frac{\left(m_{d,LR}^2\right)_{ij}}
{\sqrt{\left(m_{d,LL}^{2}\right)_{ii}\left(m^{2}_{d,RR}\right)_{jj}}},
\label{GA:dels1}
\\
\left(\drl\right)_{ij}=&
\frac{\left(m_{d,RL}^2\right)_{ij}}
{\sqrt{\left(m_{d,RR}^{2}\right)_{ii}\left(m^{2}_{d,LL}\right)_{jj}}},
&\left(\drr\right)_{ij}=
&\frac{\left(m_{d,RR}^2\right)_{ij}}
{\sqrt{\left(m_{d,RR}^{2}\right)_{ii}\left(m^{2}_{d,RR}\right)_{jj}}},
\label{GA:dels2}
\end{align}
where $i,j=1,2,3$ and $m_{d,XY}^2$ (with $X,Y=L,R$) are related to
the soft terms that appear in the soft SUSY breaking Lagrangian via
unitary transformations that transform the quark fields
from the interaction basis into the so--called 
physical super--CKM basis basis where the 
appropriate mass terms are flavour diagonal (for more details
see~\cite{OR2:bsg,FOR:bdec}). 
In the limit of MFV all $\delta_{XY}=0$.
As we are primarily concerned with flavour violation between
the third and second generations we shall, henceforth, use
the convenient shorthand $\dll=\left(\dll\right)_{23}$ and
so on.

For the remaining parameters used in our numerical analysis
we shall employ a similar parameterisation to~\cite{FOR:bdec} and
treat the soft terms defined in the physical super--CKM basis as
input. For the diagonal elements we set
\begin{align}
\left(m_{d,LL}^2\right)_{ii}&=\msq^2\;\delta_{ii},
&\left(m_{d,RR}^2\right)_{ii}&=\msq^2\;\delta_{ii},
&\left(m_{d,LR}^2\right)_{ii}&=A_d\left(m_d\right)_{ii},
\end{align}
while the remaining off--diagonal elements are related to the parameters
$\delta_{XY}^d$ via the relations defined in~\nqs{GA:dels1}{GA:dels2}.
The soft terms in the up squark sector are defined analogously.
As inputs for the Higgs sector we take $m_A$ (the mass
of the pseudoscalar Higgs), $\mu$ and $\tanb$ and use {\it FeynHiggs
2.2}~\cite{FH} to determine the remaining parameters. For the majority
of this paper we will only vary one $\dxy$ at a time unless stated
otherwise. Finally, the gaugino soft terms $M_1$ and $M_2$ are related
to the gluino mass via the usual unification relation.

\section{Limits from~\boldmath{$\bsg$}}
\label{CUR:bsg}

\FIGURE[t!]{
  \begin{tabular}{c c}
    \includegraphics[width=0.43\textwidth]{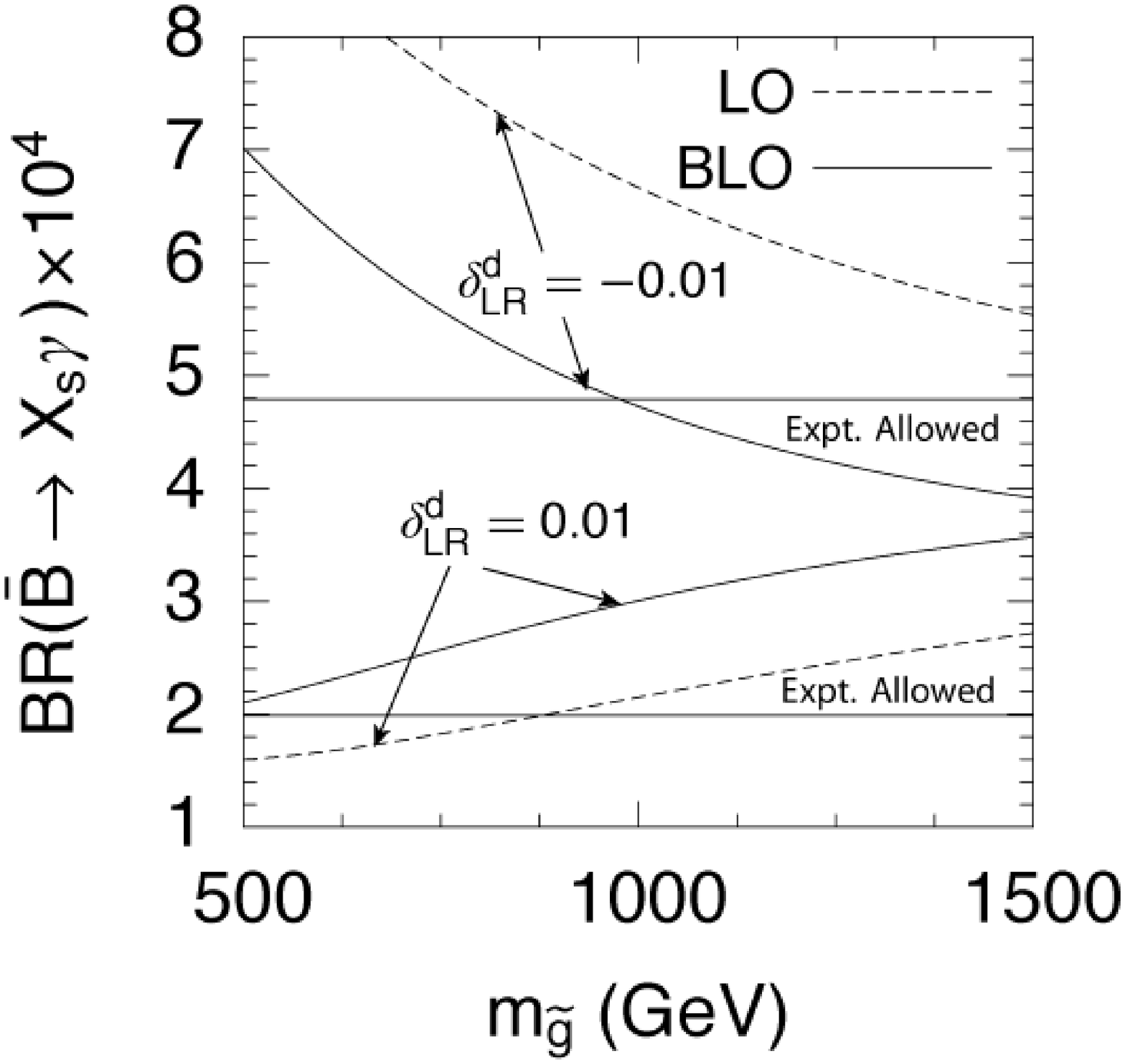}
    & \includegraphics[width=0.47\textwidth]{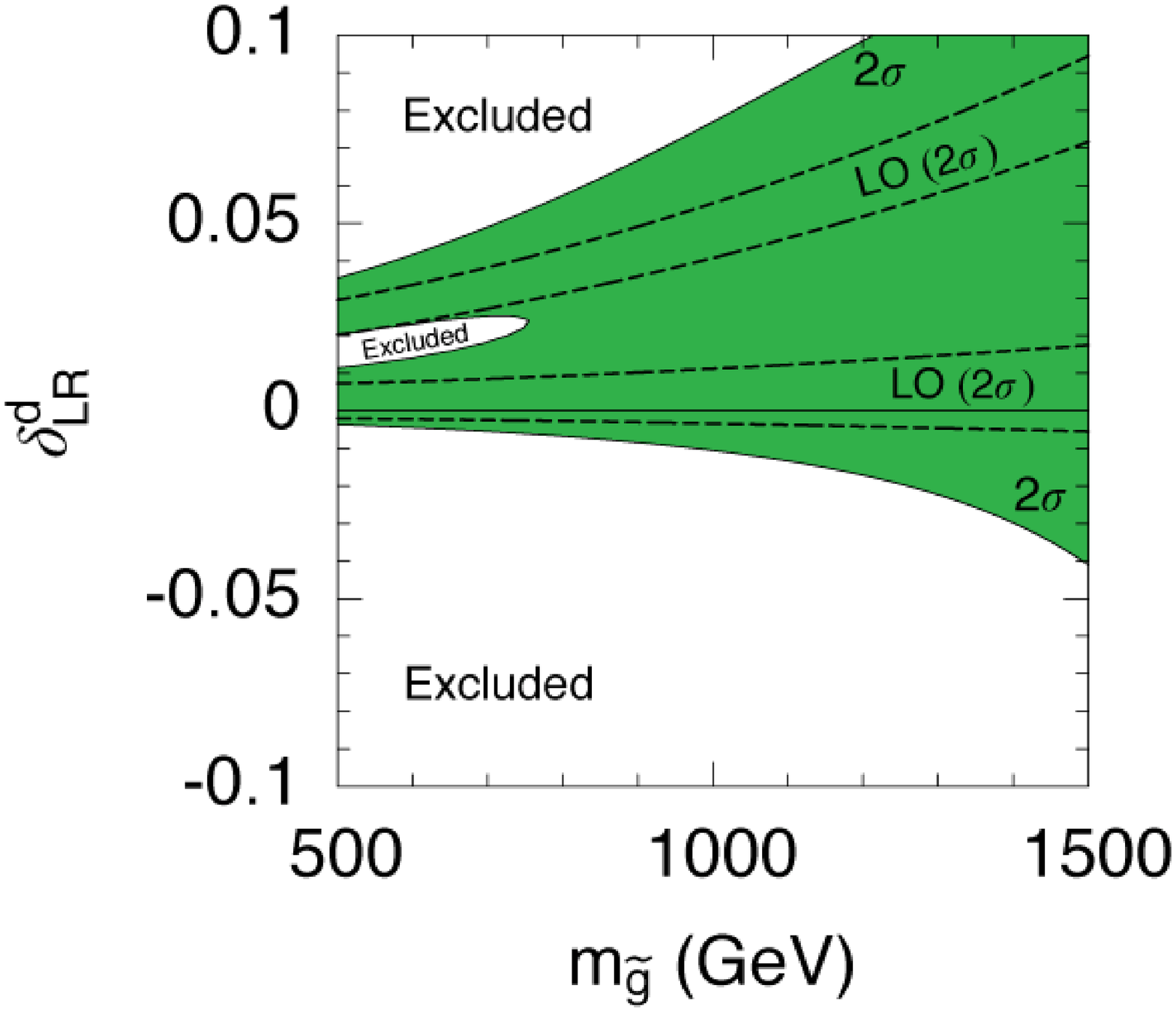}
  \end{tabular}
  \caption
{Plots depicting the effect of varying $\mgl$ on $\brbsg$
(the left panel) and the constraints from $\brbsg$ on the insertion
$\dlr$ (the right panel). In both panels the experimentally allowed
range of $\brbsg$ is taken at $2\sigma$.  In the right panel the
allowed regions corresponding to a BLO calculation are shaded in dark
green (dark grey) and bounded by solid lines.  The equivalent regions
corresponding to LO calculation is bounded by dashed lines with no
additional shading. In both panels the soft sector is parameterised as
follows $\msq=1\tev$, $m_A=\mu=-A_u=500\gev$ and $\tanb=40$.\label{bsg:mgl}}
}
The first process we shall consider is $\bsg$. Before discussing
the impact of the current constraint on all four insertions, let
us first revisit the focusing effect~\cite{OR1:bsg} that exists for the decay
in the large $\tanb$ regime. Such a situation is illustrated
in~\fig{bsg:mgl} where we show the results of a LO and a BLO analysis
beside one another for the insertion $\dlr$. As is clear
from the plots, the difference between the two calculations
tends to be rather large. In the left panel we show
the variation of the branching ratio on $\mgl$ for fixed
$\dlr$. As is evident from the figure, 
for positive and negative $\dlr$ the BLO calculation leads
to values of the branching ratio that are far more compatible with the
current experimental and SM results. As discussed
in~\cite{OR1:bsg,OR2:bsg,FOR:bdec}, this is mainly due to
a cancellation that occurs between the gluino contribution
to the decay (that tends to be reduced by BLO corrections)
and the BLO corrections to the chargino contribution.
This cancellation substantially reduces the SUSY
contribution to the decay and leads to the large reductions in the
lower limit on $\mgl$ evident in the figure. 

The implications the focusing effect has on the limits that can be placed
on the insertion are illustrated in the right panel, where
we show the $2\sigma$ contours in agreement with the experimental
result for~$\brbsg$~\eqref{brbsg:expt}\footnote{To determine the contours
we add the SM and experimental errors in quadrature and linearly
add an additional error of $0.2\times 10^{-4}$ to represent the SUSY aspect of our calculation.}
in the $\dlr-\mgl$ plane for a LO and a BLO analysis. As is evident
from the figure the difference between the two calculations can be large
and the bounds placed on the insertion can be relaxed
substantially.

\FIGURE[t!]{
  \begin{tabular}{c c}
    \includegraphics[width=0.43\textwidth]{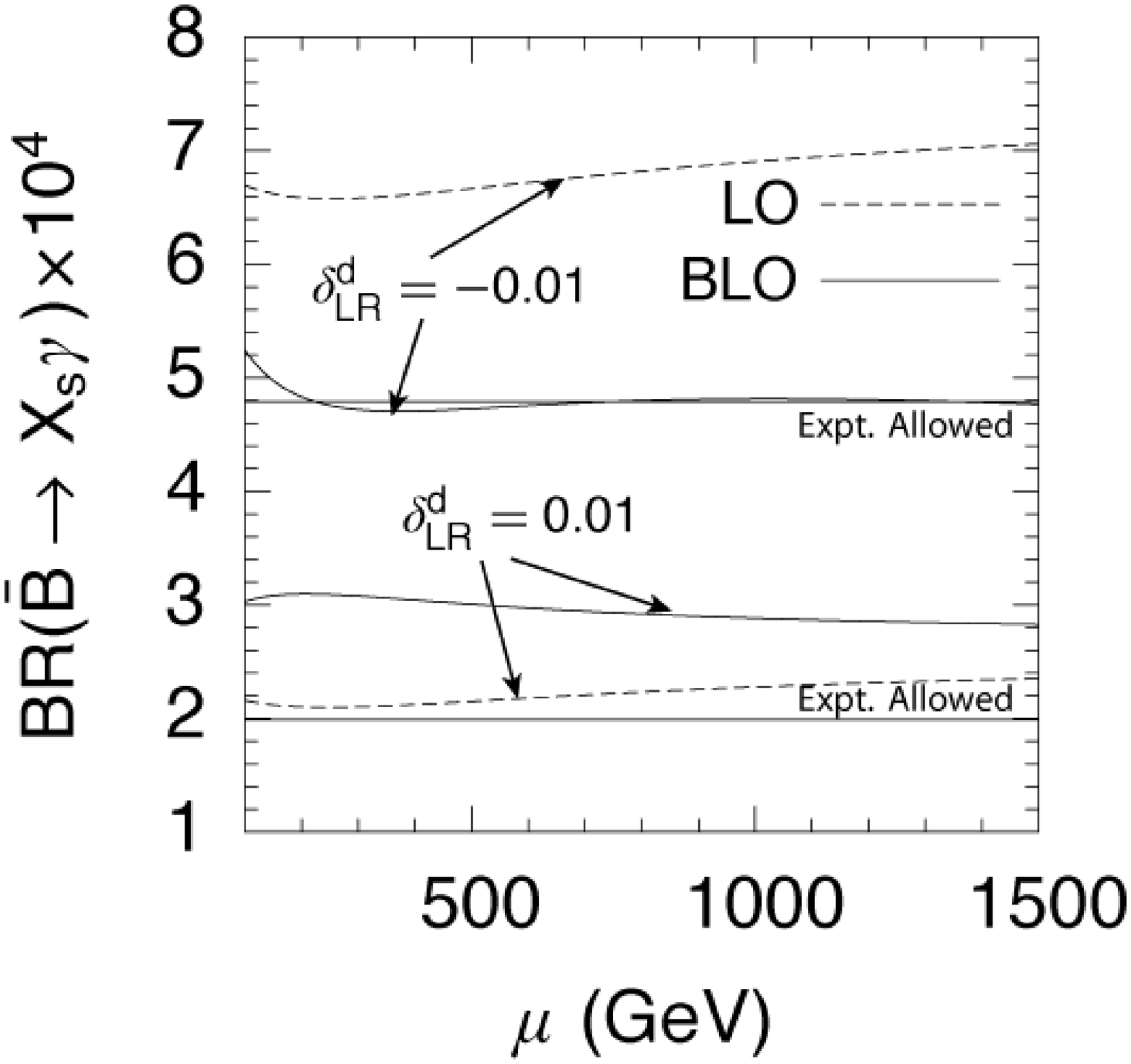}
    & \includegraphics[width=0.47\textwidth]{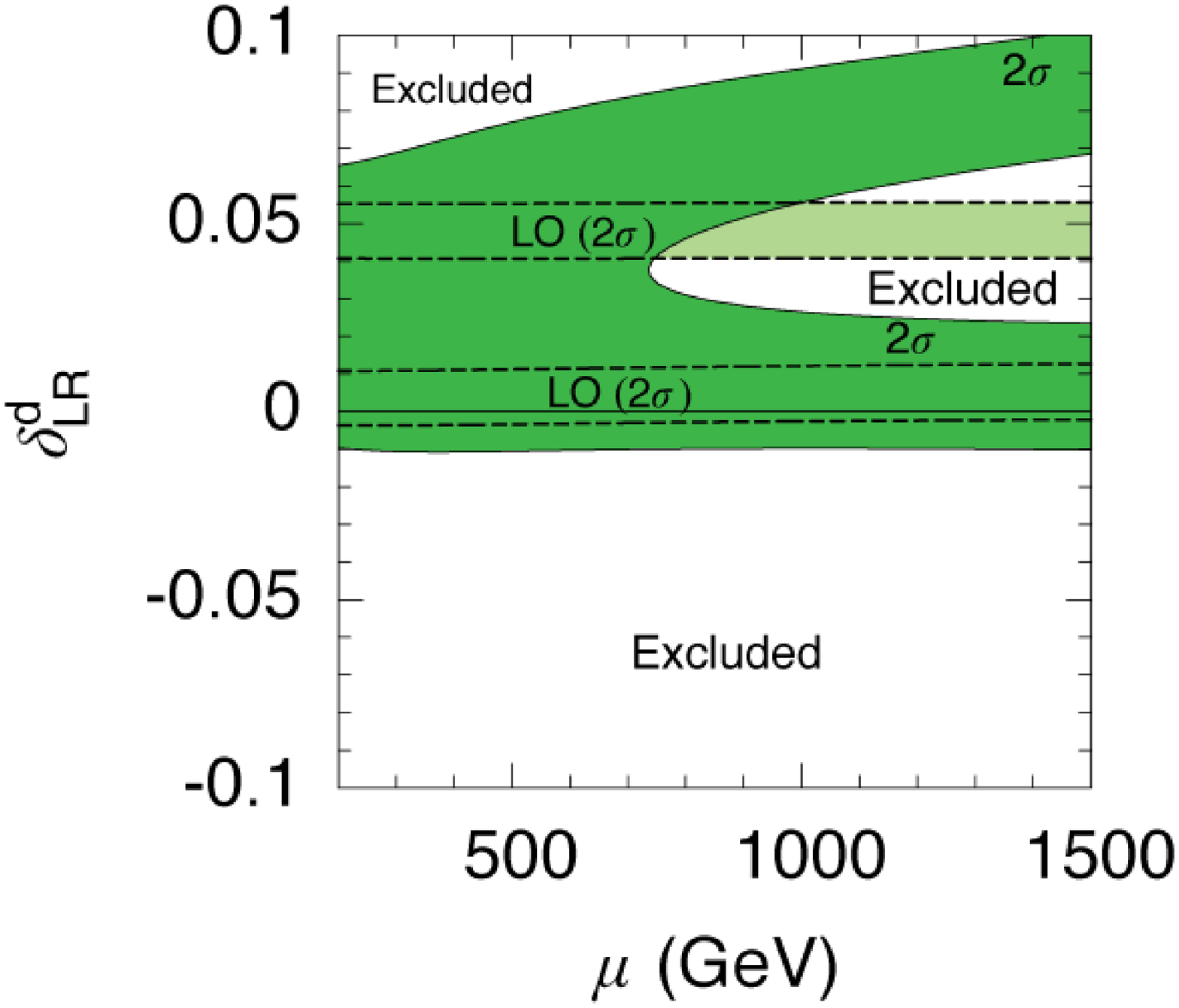}
  \end{tabular}
  \caption
  {Plots depicting the effect of varying $\mu$ on the
  constraint supplied by $\bsg$. In both panels the experimentally
  allowed range of $\brbsg$ is taken at $2\sigma$. In the left panel
  the branching ratio for the decay is plotted against $\mu$. In the
  right panel, contours corresponding to the $2\sigma$ limits on the
  branching ratio for BLO and LO calculations are shown. 
  The regions for a BLO calculation in agreement (at $2\sigma$) with the
  experimentally allowed range of $\brbsg$ are shaded in dark green
  (dark grey) and bounded by solid lines. The equivalent regions for a
  LO calculation are bounded by dashed lines with no additional
  shading when they coincide with the allowed region corresponding to
  a BLO calculation and by light green (light grey) otherwise. In both
  panels the soft sector is parameterised as follows
  $\msq=\mgl=1\tev$, $m_A=-A_u=500\gev$ and $\tanb=40$.\label{bsg:alt}}
}
It is natural to ask how, for fixed $\msq$ and $\mgl$, the
remaining free parameters affect the focusing effect at large
$\tanb$. The largest contribution to the focusing effect
usually arises from the cancellation that takes
place between the gluino and BLO chargino contributions
to the decay. Regions of parameter space that
maximise the chargino contribution to the decay with
respect to the gluino contribution would therefore
be expected to lead to regions where the focusing
effect is greatest. As the squark masses essentially
enter the gluino and chargino contributions to
the decay, the only difference between the two stems
from the appearance of the chargino masses rather
than the gluino mass in the relevant matching conditions.
The focusing effect is therefore usually maximised when
the two masses that enter the diagonal terms of the
chargino mass matrix, namely $M_2$ and $\mu$ are
much smaller than $\mgl$. While the soft mass $M_2$ is typically
lighter than the gluino mass in most SUSY breaking models,
the $\mu$ term on the other hand can be much more variable.
One should note that whilst increasing $\mu$ will often
decrease the BLO chargino correction, it tends to similarly
decrease the gluino contribution to the decay
through its dependence on the BLO correction $\epsilon_s$
(for the relevant expression see Eq.~(4.15) in~\cite{FOR:bdec}). This
compensation tends to reduce the overall $\mu$ dependence of the
focusing effect.

The $\mu$ dependence of the branching ratio
is presented in the left panel in~\fig{bsg:alt}.
As shown in the discussion above, increasing $\mu$ 
tends to have relatively little impact on the branching
ratio. This is caused by the interplay between the BLO corrections
to the gluino and chargino contributions mentioned earlier.
Similar behaviour is exhibited for RL and RR insertions.
For LL insertions, on the other hand, the $\mu$ dependence
is slightly more variable as a chargino mediated contribution
exists at LO (unlike the other three insertions).
Turning to the right panel, we show
the bounds on the insertion $\dlr$ and how they vary
with $\mu$. As is evident from the plot, in contrast to the
right panel of~\fig{bsg:mgl}, the bounds on the insertion
tend to become more stringent with incerasing $\mu$. However, we should point out
that, even in regions of parameter space where $\mu>\mgl$,
the differences between BLO and LO calculations can
still be large. 

\FIGURE[t!]{
  \begin{tabular}{c c}
    \includegraphics[width=0.49\textwidth]{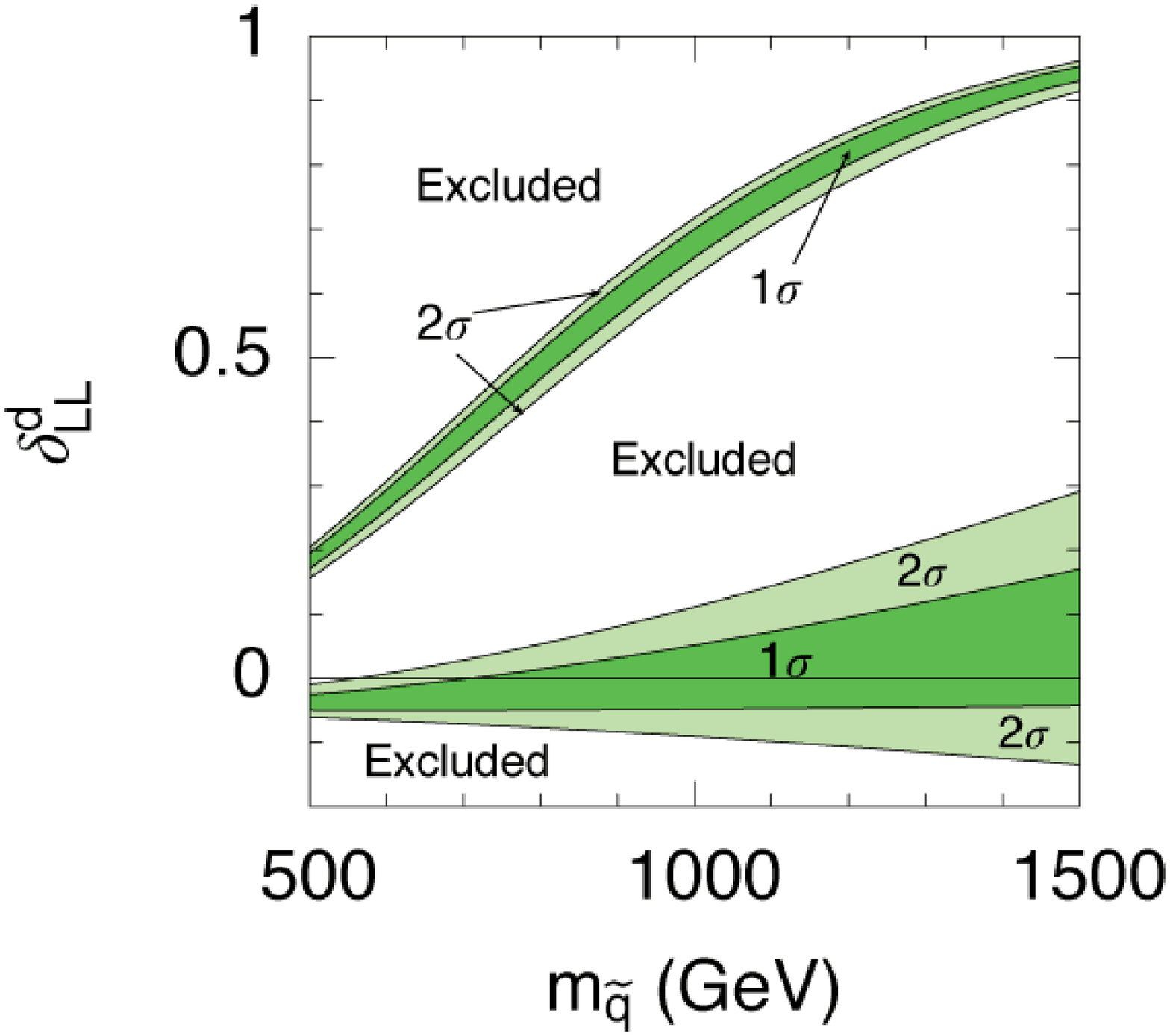}
    & \includegraphics[width=0.49\textwidth]{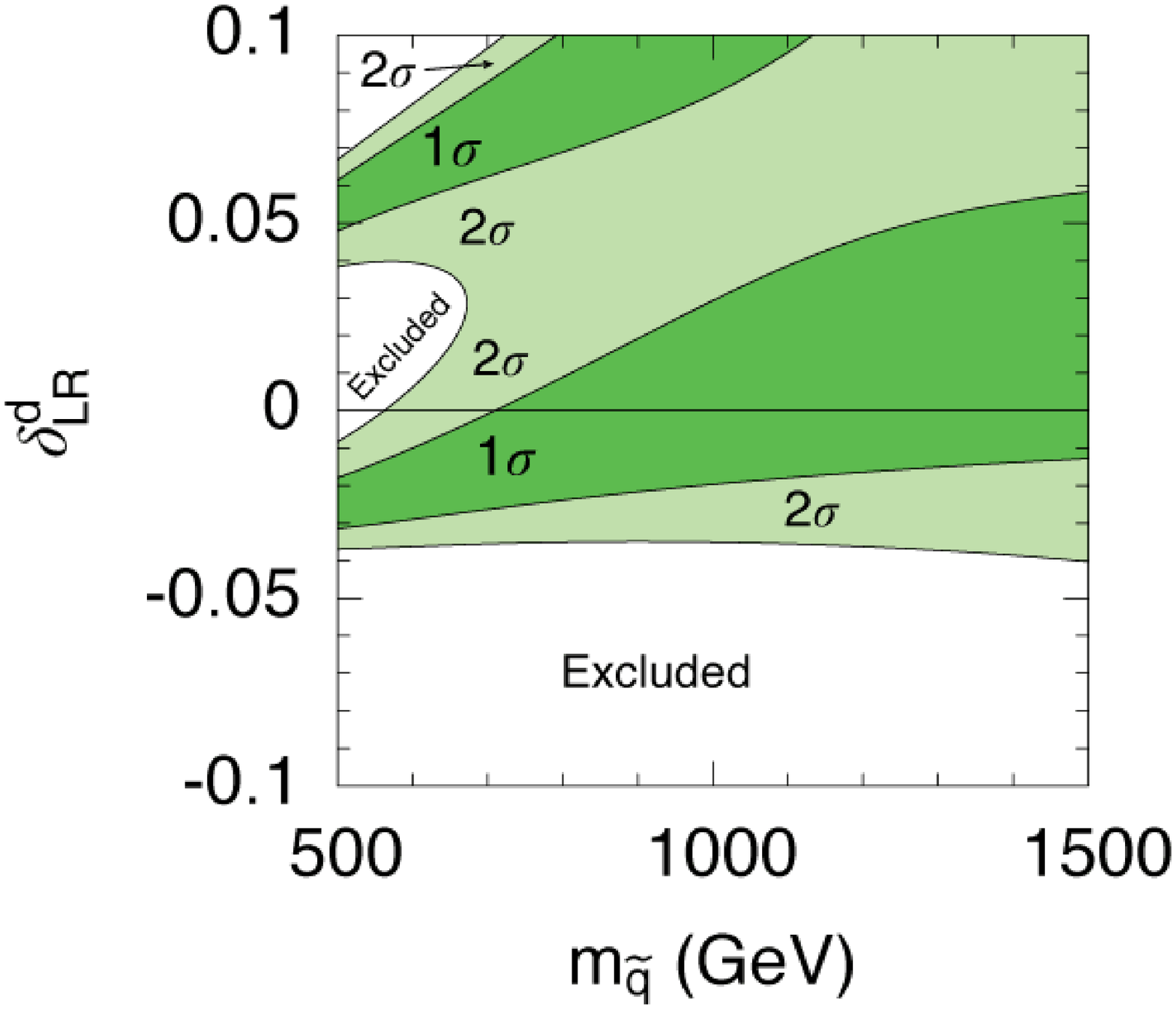}\\
    \includegraphics[width=0.49\textwidth]{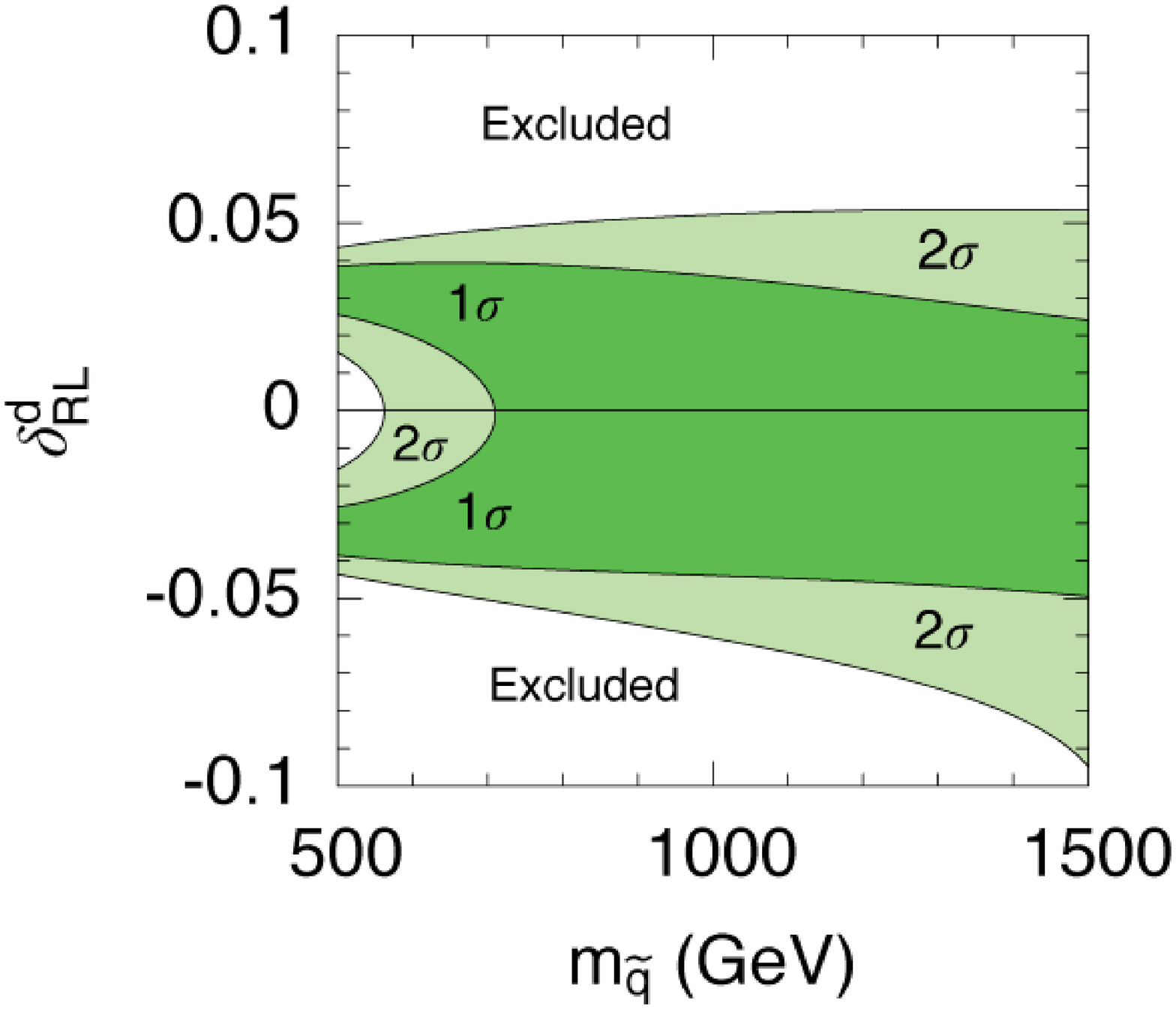}
    & \includegraphics[width=0.49\textwidth]{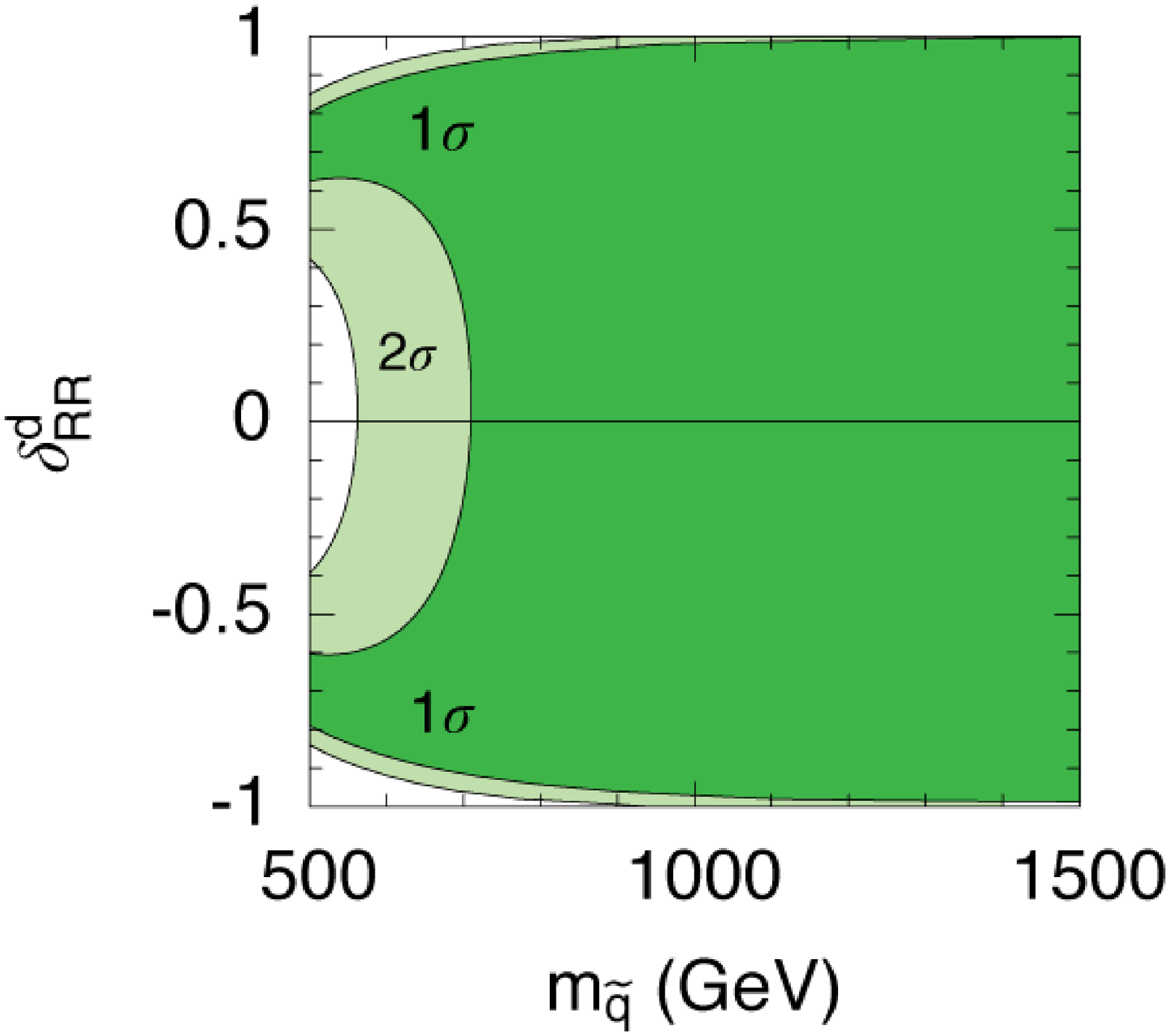}
  \end{tabular}
  \caption
  {Contour plots
  depicting the impact of $\brbsg$ on the $\msq$---$\dxy$ plane. In each
  plot the soft sector is described as follows $\mgl=\sqrt{2}\msq$,
  $A_u=-\msq$, $\mu=\msq/\sqrt{2}$, $m_A=500\gev$ and $\tanb=40$. The
  light green (light grey) regions highlight areas that agree within
  $2\sigma$ with the experimental result, the dark green (dark grey)
  regions within $1\sigma$.\label{bsg:cnt}}
}
The bounds on each flavour violating parameter imposed by the
decay at large $\tanb$ are illustrated in~\fig{bsg:cnt}.
In the top left panel, depicting the limits on the insertion $\dll$,
we can see that the bounds are fairly
strict. This is because, at leading order, two large contributions
to the decay appear. The first is associated with gluino
exchange, where the dominant contribution stems from the $\tanb$
enhanced correction that appears at second order in the
MIA. The second is associated with chargino
exchange. As flavour violation between left handed squarks
in the up and down sectors is related by $\sutwo$ symmetry, it
is likely that, if $\dll$ is non--zero, the corresponding measure
in the up--squark sector -- $\ull$ -- is non--zero as well. If
$\ull$ is non--zero, it is possible to induce a LO contribution
via chargino exchange that benefits from appearing at first
order in the MIA as well as being $\tanb$ enhanced
(for a complete set of analytic expressions see~\cite{FOR:bdec}). 
The presence of this LO chargino contribution tends to negate
a large part of the focusing effect as it interferes with the
cancellation between the BLO part of the chargino contribution, 
that is proportional to $\dll$, and the LO gluino contribution. 
As such, even once BLO corrections are taken into account, the
impact of the $\bsg$ constraint on LL insertions at large
$\tanb$ tends to be more stringent than the other
three insertions.

In the top--right panel the bounds on the insertion $\dlr$ are
illustrated. As is evident from the plot, the bounds on the insertion
are fairly weak. This is primarily due to the focusing
effect discussed earlier in this section. At LO
the constraints on the insertion $\dlr$ are more
strict (see, for example, the plot on the right of~\fig{bsg:mgl})
due to the chiral enhancement $\mgl/m_b$ the leading
order contribution receives. However, once one includes the
dominant effects that appear BLO, cancellations between the LO
gluino contribution and the BLO corrections to the chargino
contribution can lead to a significant relaxation of the bounds
on the insertion. 

The lower--left and right panels illustrate the bounds imposed on the
insertions $\drl$ and $\drr$. As is evident from the plots the
bounds derived from $\bsg$ on these two insertions are rather
slight. This is due to the fact that the contributions due to the
two insertions cannot interfere with the SM contribution as the
dominant contributions due to the insertions are to the primed
Wilson coefficients. The bounds on each insertion are therefore
fairly symmetric. For the insertion $\drl$ the BLO correction to
the charged Higgs vertex arising from higgsino exchange leads to
slight deviations from this behaviour~\cite{FOR:bdec}.

As an aside let us briefly discuss the decoupled limit
$\msq,\mgl\to\infty$ where we are effectively left with
a two Higgs doublet model that includes the $\tanb$ enhanced
threshold corrections discussed in detail in~\cite{FOR:bdec}. In this
limit the insertions $\dlr$ and $\drl$, which scale as $1/M_{SUSY}$,
are naturally tiny, and the only meaningful constraint
$\bsg$ can place on sources of SUSY flavour violation is
on the insertions $\dll$ and $\drr$. 
For the insertion $\dll$, the threshold corrections to the charged
Higgs vertex can lead to useful bounds
that can compliment the constraints supplied by 
$\bsm$ and $\bbb$ mixing rather well. For the insertion $\drr$, on
the other hand, the corrections to the charged Higgs vertex
only affect the primed Wilson coefficients and the constraints
that can be imposed on the insertion are rather mild to say the least.
We shall discuss this issue further at the end of section~\ref{CUR:LimSing}.

Finally, let us discuss the scenario $\mu<0$.
Generally in MFV the $\bsg$ constraint
for $\mu<0$ tends to be fairly stringent as the chargino
and charged Higgs contributions interfere constructively with
the SM result. In the GFM scenario, on the other hand, this situation
can be avoided by inducing relatively mild GFM corrections
that interfere destructively with the other SUSY contributions.
For $\mu<0$ it is therefore possible to reconcile
SUSY contributions with experiment for
either negative LL insertions or positive LR insertions. 
As the negative $\mu$ case was discussed in~\cite{OR1:bsg,OR2:bsg}
(see, for example, Fig. 12 of~\cite{OR2:bsg})
we shall not cover it in detail here, however, we shall
return to it later when we supplement the $\bsg$ constraint
with those provided by $\bsm$ and $\bbb$ mixing.

\section{Limits from~\boldmath{$\bsm$}}
\label{CUR:bsm}

\FIGURE[t!]{
  \begin{tabular}{c c}
    \includegraphics[width=0.49\textwidth]{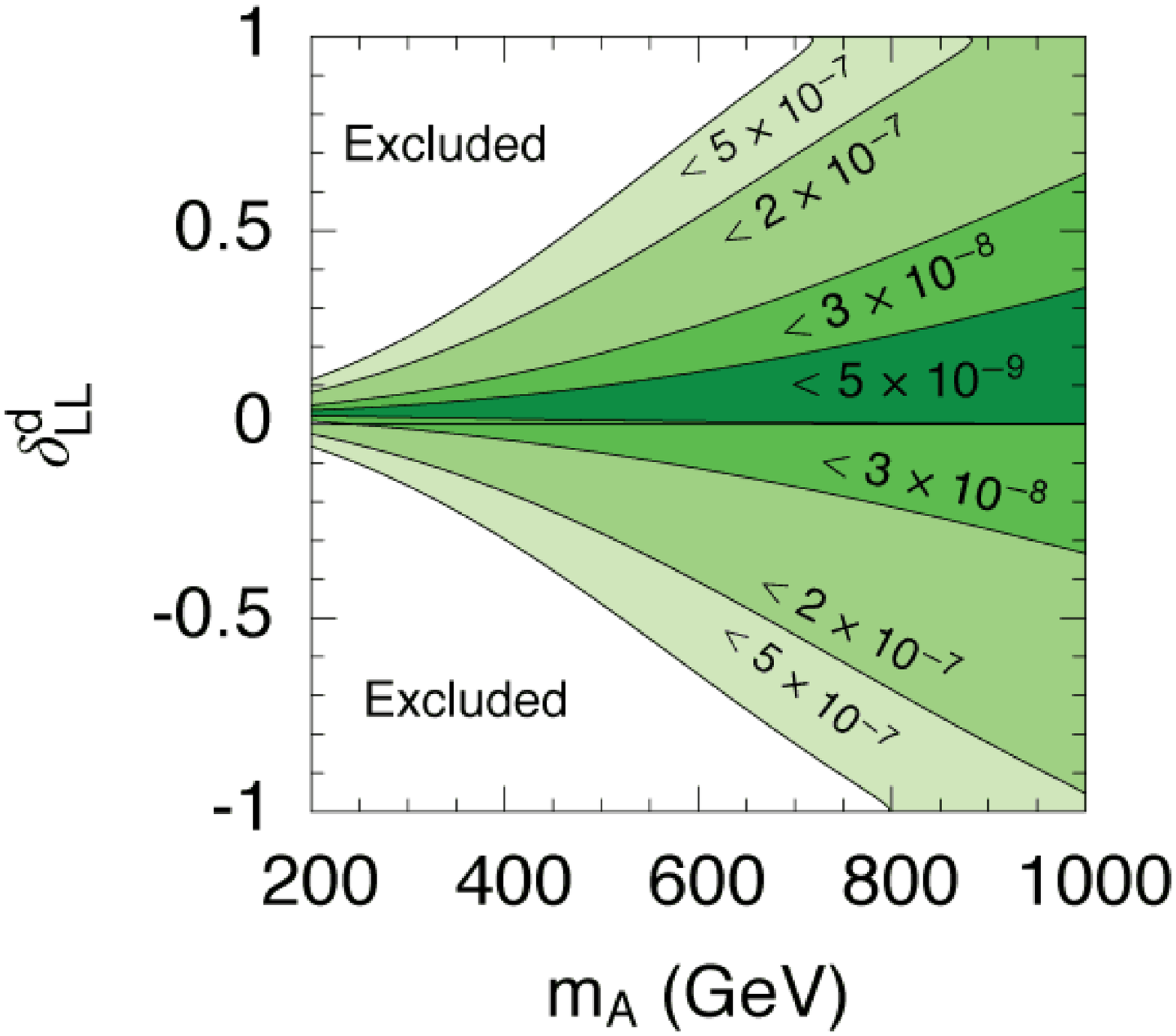}
    & \includegraphics[width=0.49\textwidth]{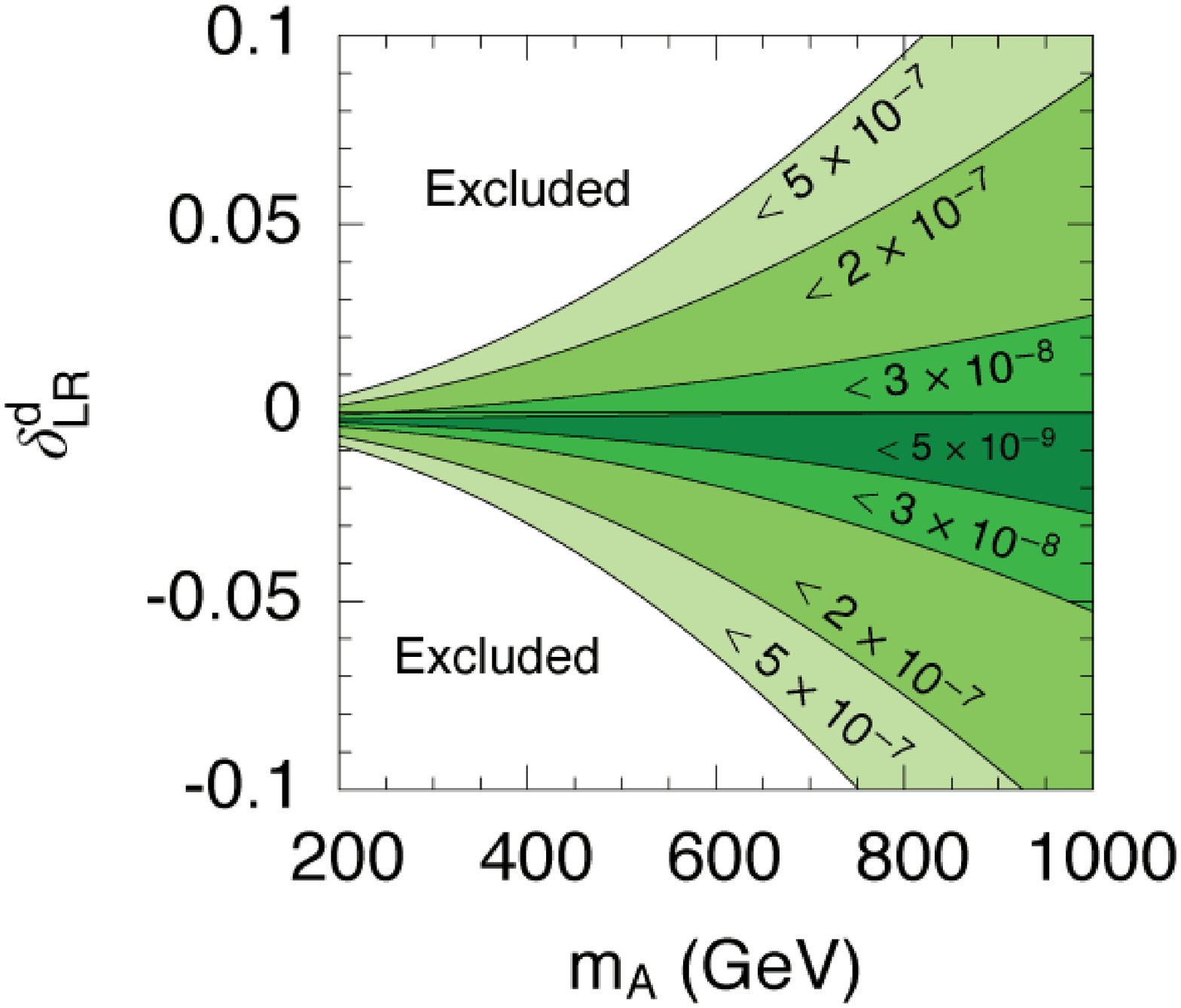}\\
    \includegraphics[width=0.49\textwidth]{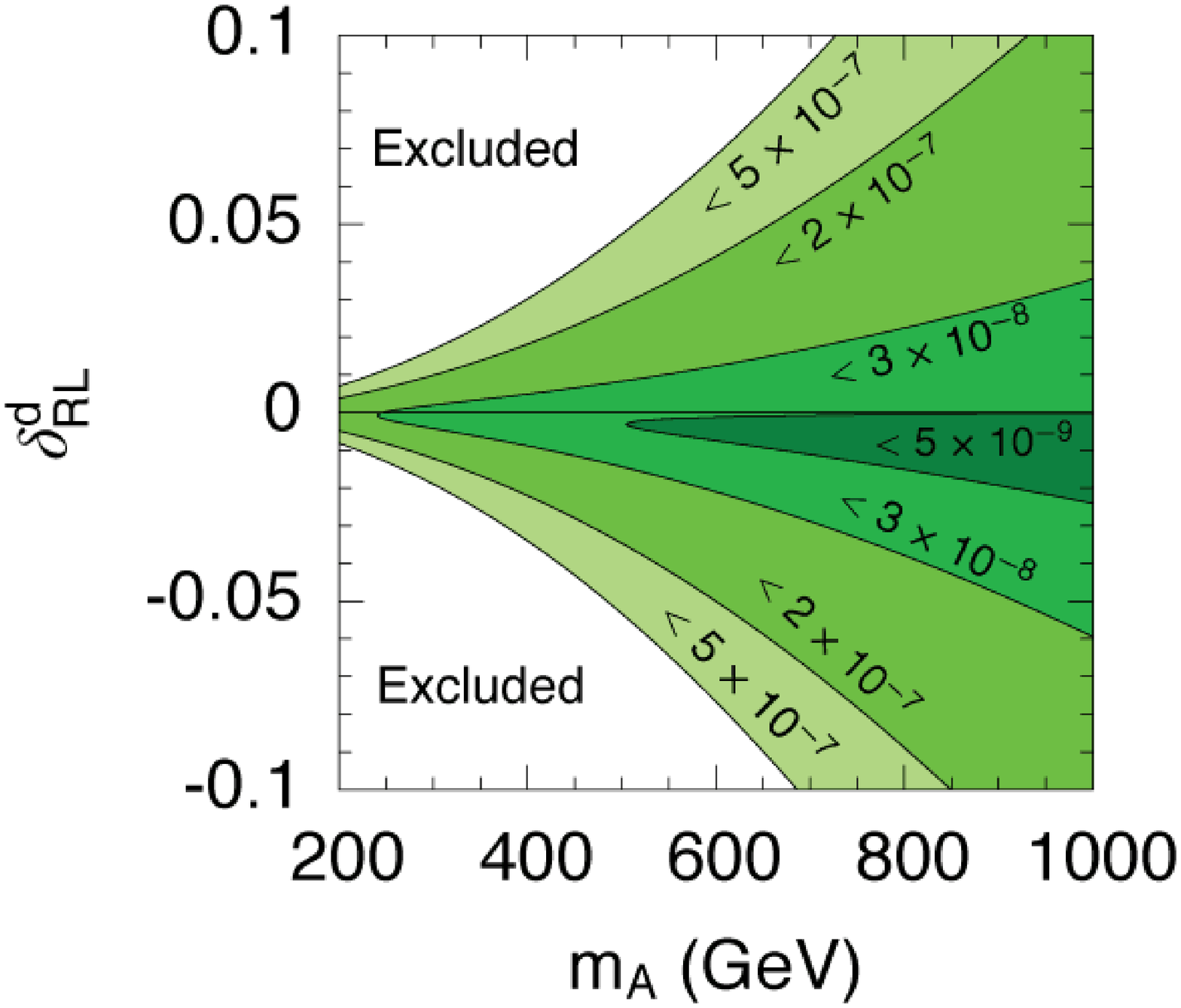}
    & \includegraphics[width=0.49\textwidth]{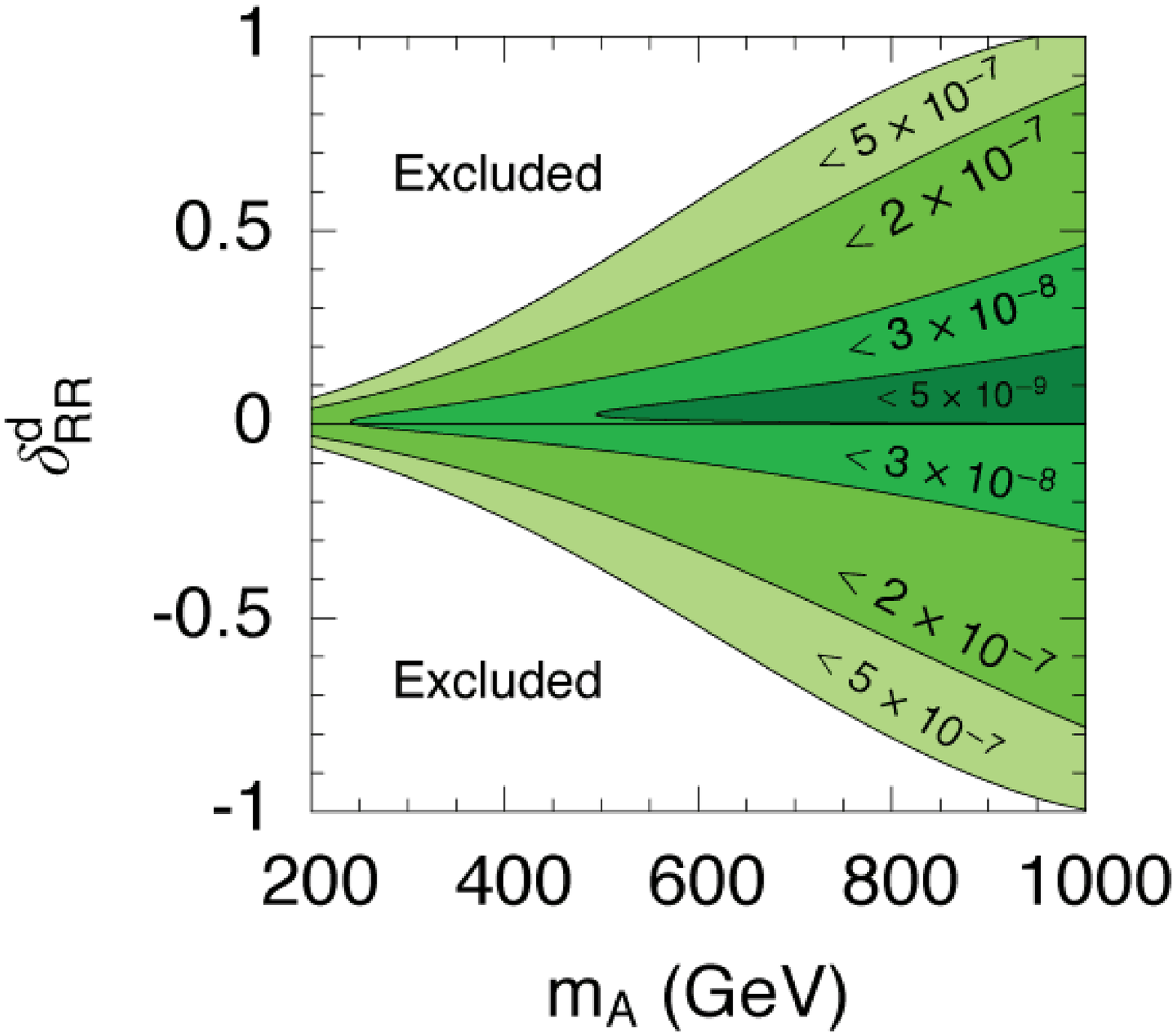}
  \end{tabular}
  \caption
  {Contour plot depicting the regions of parameter space
  excluded by increasing the limit on $\brbsm$ for $\tanb=40$. 
  The various shaded regions correspond to regions that satisfy
  the following upper limits: $5\times 10^{-7}$ light green (light
  grey); $2\times 10^{-7}$ light--medium green (light--medium grey);
  $3\times 10^{-8}$ dark--medium green (dark--medium grey) and
  $5\times 10^{-9}$ dark green (dark grey). The soft sector is
  parameterised as follows $\msq=\mgl=1\tev$ and $\mu=-A_u=500\gev$.
  \label{bsm:cnt}}
}
It is well known that SUSY threshold corrections to the neutral
Higgs vertex can lead the branching ratio for the decay
$\bsm$ to vary as $\tan^6\beta$~\cite{NH:th,NH:th2,NH:rev}.
Coupled with a relatively weak dependence on the SUSY mass scale
(the branching ratio varies as $1/m_A^4$ rather than $1/M_{SUSY}^4$),
the constraint supplied by the Tevatron upper bounds on the
decay can prove to be useful even when the coloured SUSY
particles (\ie~the squarks and gluino) are heavy. 

Figure~\ref{bsm:cnt} illustrates the bounds imposed by the decay on each
of the insertions for $\tev$ scale squark masses and varying
pseudoscalar Higgs mass in the large $\tanb$ regime. All four
plots illustrate the useful bounds that can be placed on SUSY
flavour violation using the decay, if $m_A$ is relatively small
(compared to the mass of the squarks).

Before discussing the other aspects of the figures, let us briefly
provide a rough recipe that allows those investigating flavour violating
effects in the squark sector to decide whether the $\bsm$
constraint is relevant to their study or not:
\begin{align}
\lvert\dll\rvert&\approx 0.35\left(\frac{40}{\tanb}\right)^3\left(\frac{500\,\gev}{m_A}\right)^{-2};
&
\lvert\dlr\rvert&\approx 0.026\left(\frac{40}{\tanb}\right)^3\left(\frac{500\,\gev}{m_A}\right)^{-2};
\\
\lvert\drl\rvert&\approx 0.030\left(\frac{40}{\tanb}\right)^3\left(\frac{500\,\gev}{m_A}\right)^{-2};
&
\lvert\drr\rvert&\approx 0.26\left(\frac{40}{\tanb}\right)^3\left(\frac{500\,\gev}{m_A}\right)^{-2}.
\end{align}
Let us note that these formulae are highly approximate and
serve only as a rough guide and should not replace the limits
derived from a full numerical analysis.

The effect of improving the limit on $\bsm$ is also shown in
Figure~\ref{bsm:cnt}. It can be seen from all four plots that even
the transition from the published D\O~result~(\ref{bsmm:d0:publ})
to the preliminary CDF~(\ref{bsmm:cdf:prel}) 
result reduces the allowed region in the
GFM parameter space by a sizeable amount. It is also
apparent from the figure that an improvement of the limit to
roughly $3\times 10^{-8}$ (a limit achievable at the Tevatron)
will provide an excellent constraint
on GFM in the large $\tanb$ regime. Finally, let us point
out the cancellations that can occur for each insertion. For
flavour violation in the LL and LR sectors, direct interference
with the MFV contributions is possible and values for the branching
ratio approaching the SM model value are possible for $m_A$ as low
as $200\gev$, or less. For flavour violation in the RL and RR sectors,
on the other hand, the cancellations that can occur are more complicated.
In general, the insertions act to increase the branching ratio.
However, at large $m_A$ regions comparable to the SM can become
viable. These regions appear due to a cancellation, between the
Wilson coefficients $C_P^{\prime}$ and $C_{10}$, that arises
when one calculates the branching ratio for the decay.
It should be noted that in all four panels the MFV contribution
(\ie~all $\dxy=0$) is such that SM--like values of the branching
ratio are impossible. This is, of course, due to the choice 
of parameters we make, rather than being a general feature of the MFV
contribution at large $\tanb$.

The plots in~\fig{bsm:cnt} also serve to illustrate how 
a lower bound on the pseudoscalar mass derived
from the upper limits on BR$\left(\bsm\right)$ is sensitive to the
assumed flavour structure, 
and how any upper
limit that appears (should $\bsm$ be observed at the Tevatron for
example) might be altered beyond MFV. As an example let us
consider the top--left 
and bottom--right panels in~\fig{bsm:cnt}. The top--left panel
illustrates that one can typically avoid the lower bound
on $m_A$ altogether even for a relatively
minor amount of flavour violation in the LL sector. While
this assumption smacks of fine--tuning, one should remember
that LL insertions are by far the easiest to generate through
RG running in the MSSM (see, for example~\cite{LL:rad}),
and such notions, therefore, might not be as distasteful as one
might initially think. Beyond these cancellations that appear for
small $\dll$ it is apparent from both panels that the presence of
flavour mixings in the squark sector generally act to increase the
lower bound on $m_A$.

It was discussed in~\cite{DH:bsm} that if one assumed MFV and
imposed an upper bound on $\tanb$, it would be possible to place
some sort of upper bound on the pseudoscalar Higgs mass $m_A$. If
one allows for GFM in the squark sector it is apparent from all
four plots that this result will no longer hold (one can travel
along a given band in the plots towards arbitrarily large $m_A$)
and one would probably need to resort to a more general analysis
that takes into account a variety of other processes before
any concrete bound could be derived.

\section{Limits from~\boldmath{$\bbb$} Mixing}
\label{CUR:bbb}

\FIGURE[t!]{
  \begin{tabular}{c c}
    \includegraphics[width=0.49\textwidth]{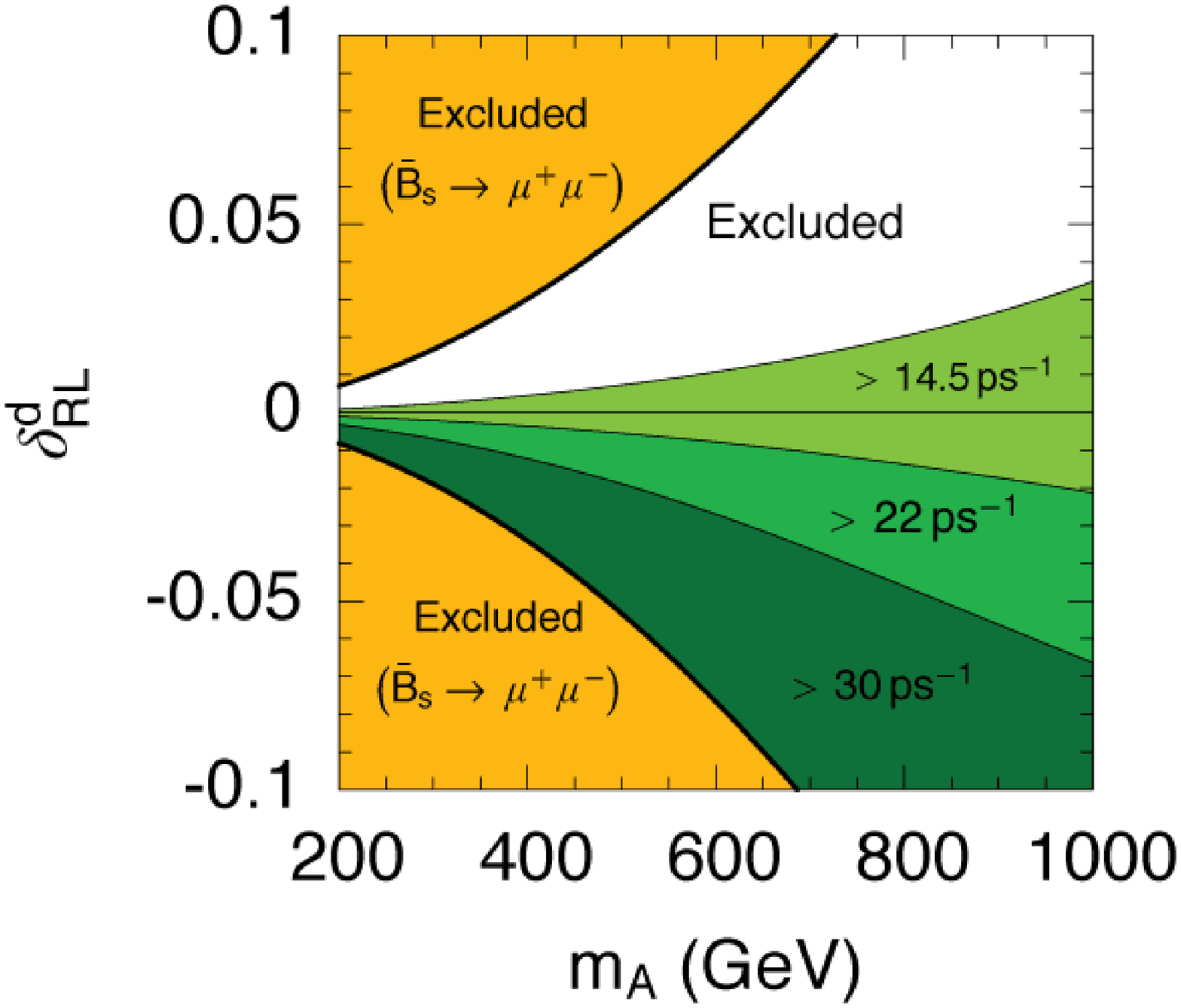}
&    \includegraphics[width=0.49\textwidth]{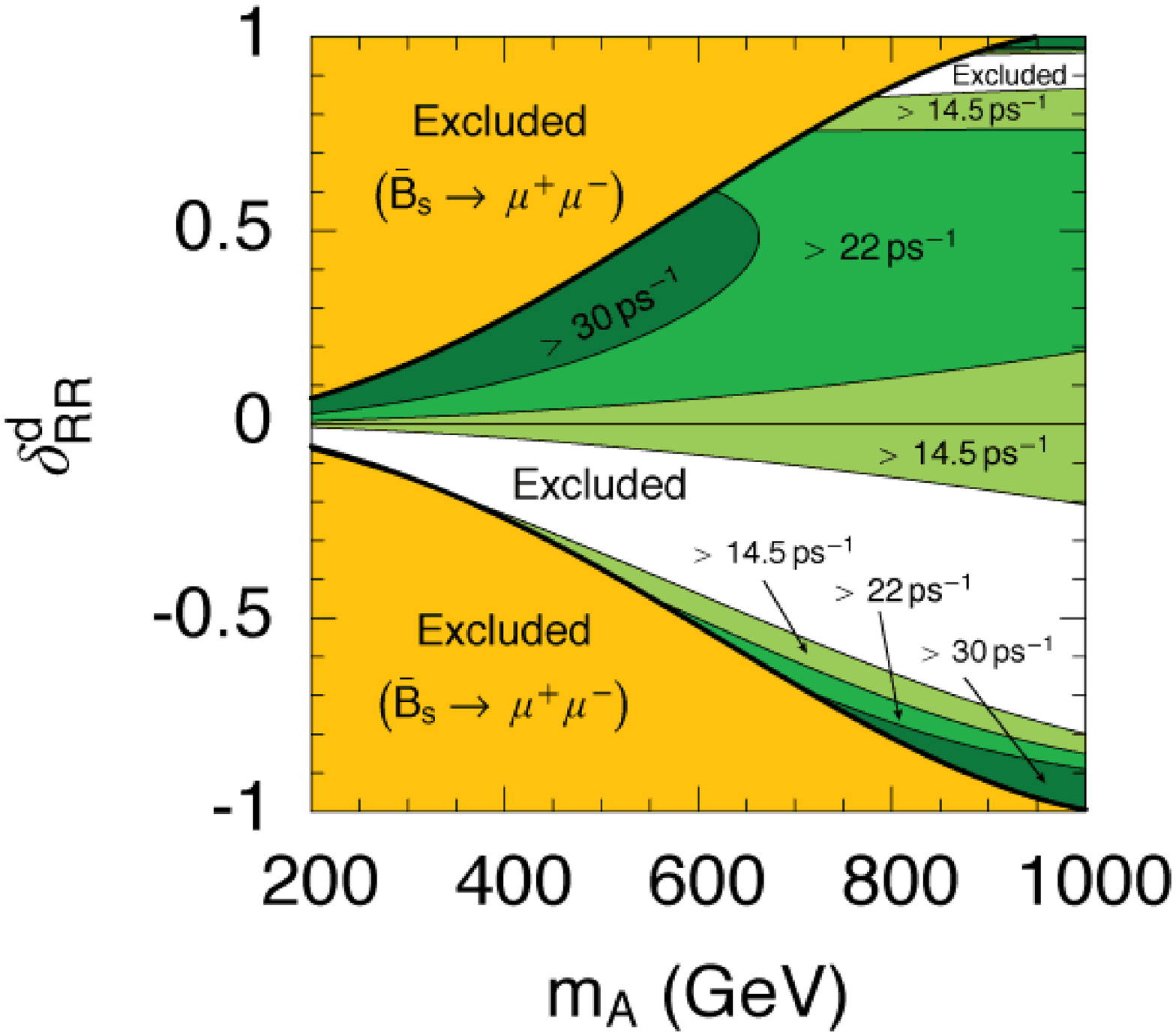}
  \end{tabular}
  \caption
  {Plots illustrating the dependence on $m_A$ and $\dxy$
  of $\delmbs$ once the published D{\O} limit
  $\brbsm<5.0\times10^{-7}$ has been 
  taken into account (the region shaded in orange/medium grey and
  bounded by a thick line). The shading is as follows: white regions
  are excluded by the lower limit  $\delmbs>14.5\ps^{-1}$;
  light green (light grey) regions indicate values of $\delmbs$ in
  the interval $14.5$--$22\ps^{-1}$; medium green (medium grey)
  $22$--$30\ps^{-1}$; dark green regions (dark grey) indicate
  points in parameter space where $\delmbs>30\ps^{-1}$. The soft
  sector is parameterised in the same manner as~\fig{bsm:cnt}.\label{bbb:cnt}}
}
Large contributions to $\bsm$ have been known for some time
now to induce similar corrections in the $\bbb$ mixing system.
These corrections arise through Feynman diagrams that feature two of
the flavour changing neutral Higgs penguin vertices that feature
in the $\tanb$ enhanced contributions to $\bsm$ (see~\cite{FOR:bdec}
for more details). In the MFV scenario such corrections typically
lead to reductions to $\delmbs$ taking it closer to the current
experimental limit~\cite{BCRS:bbm}. Since LL insertions affect the
corrected neutral Higgs vertex in a similar manner to the MFV
corrections~\cite{FOR:bdec} the overall effect is similar
to that found in MFV. The only exception is
that the effects are generally more exaggerated due to the enhancement
that the GFM correction receives due to the presence of the strong
coupling constant in the relevant matching conditions. On the other
hand, the contributions due to the insertion $\dlr$ tend to be small
and generally are bounded by the $\bsm$ constraint~\cite{FOR:bdec}.

Turning to flavour violation in the RL and RR sectors, as we
pointed out in~\cite{FOR:bdec}, the contribution
where one neutral Higgs penguin is mediated by chargino exchange and
the other by gluino exchange (thereby eliminating the suppression
by $m_s$ that blights the MFV, LL and LR contributions), can
yield large effects on $\delmbs$. Such effects
are illustrated in~\fig{bbb:cnt} where we show contours of
$\delmbs$ for varying $\drl$, $\drr$ and $m_A$ for large $\tanb$.
As is evident from both panels there is far more variation in $\delmbs$
compared to the MFV limit $\dxy=0$.
Negative values of $\drr$ and positive values $\drl$, in particular,
seem to be disfavoured by the current lower bound on $\delmbs$. 
It should be noted that the effects induced by varying $\drl$ 
arise entirely from BLO corrections and would be absent in a
purely LO calculation~\cite{FOR:bdec}. From the plots
it should also be noted
that values for $\delmbs$ far in excess of the SM prediction~\eqref{bbb:SM} are also
allowed by the current upper bound on BR$\left(\bsm\right)$.
This provides a unique signature for the existence of flavour
violation in the RR and RL sectors if $\tanb$ is large. For example,
if the Tevatron were to measure $\brbsm$ at a
level far in excess of the SM prediction, and $\delmbs$ was
subsequently measured at a value substantially larger than the SM prediction
(rather than lower as predicted by MFV~\cite{BCRS:bbm}),
such observations could only point towards the existence of
non--minimal flavour structure in either the RL or RR sectors.

\FIGURE[t!]{
  \begin{tabular}{c c}
   \includegraphics[width=0.49\textwidth]{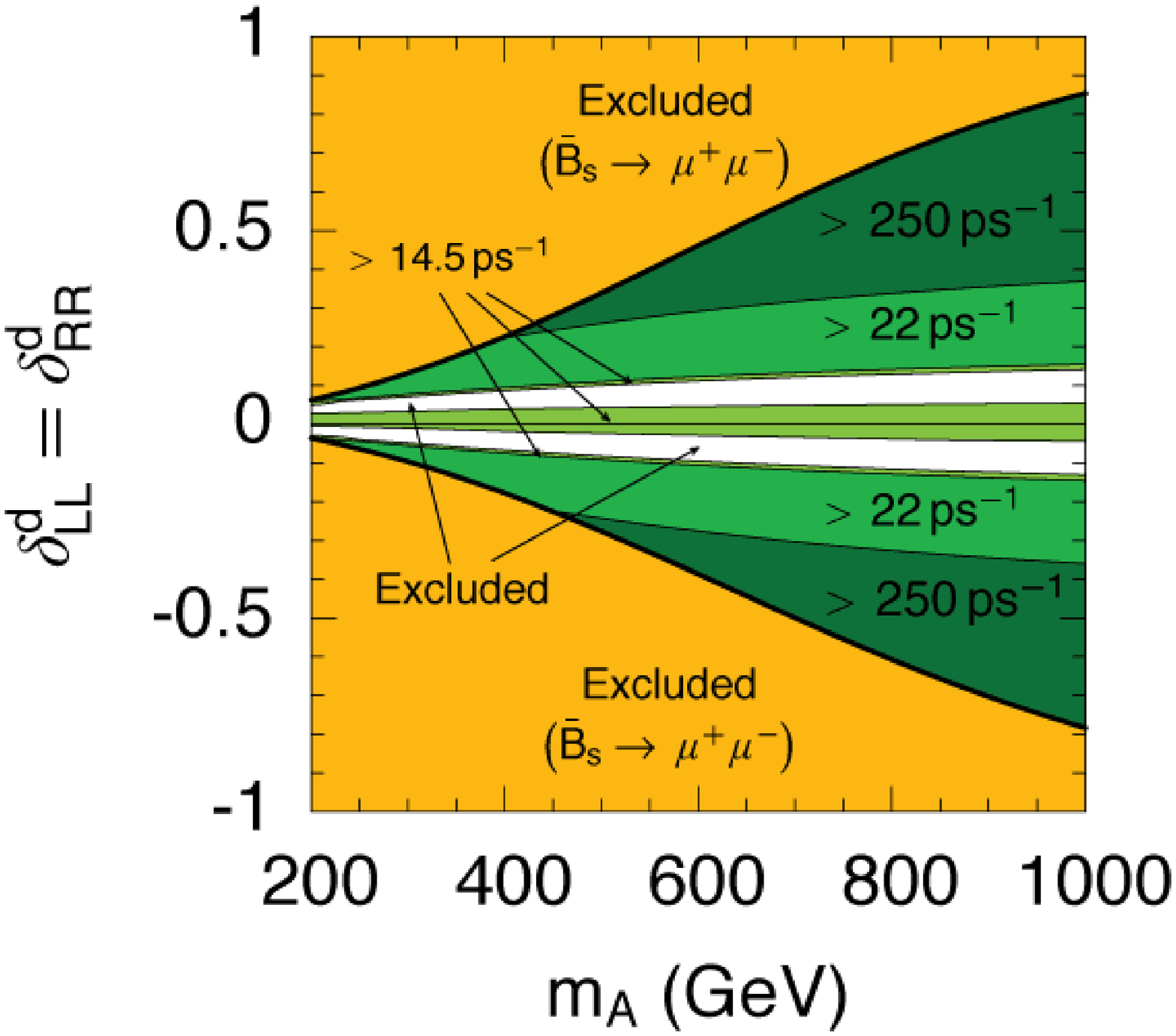}
   & \includegraphics[width=0.49\textwidth]{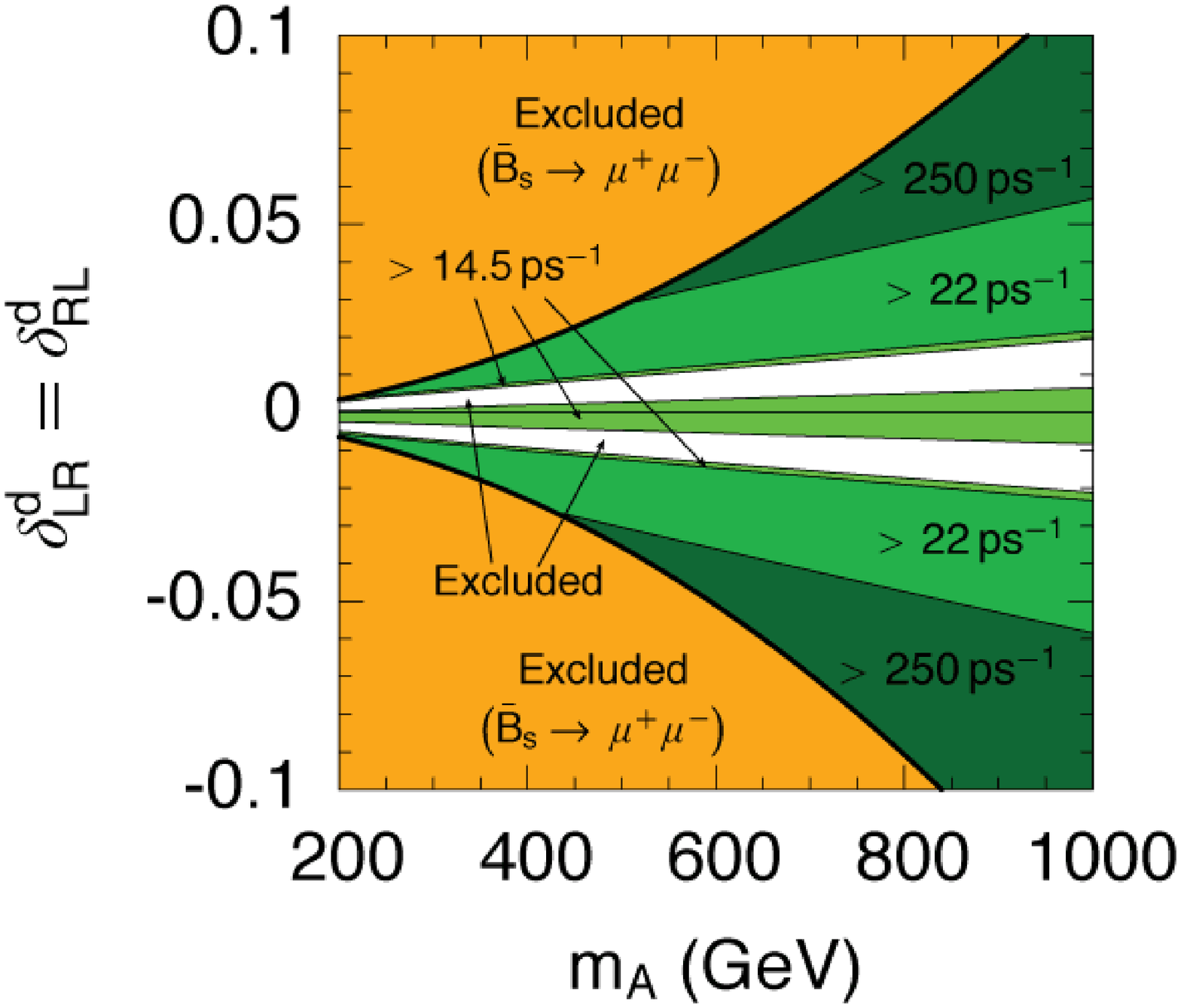}
  \end{tabular}
  \caption
  {Plots illustrating the variation of $\delmbs$ once
  the published D{\O} limit $\brbsm<5.0\times10^{-7}$ has been
  taken into account 
  (the region shaded in orange/medium grey and bounded by a thick
  line). The panels are shaded as follows: white regions are
  excluded by the lower limit $\delmbs>14.5\ps^{-1}$; light
  green (light grey) regions indicate values of $\delmbs$ in the interval
  $14.5\ps^{-1}$--$22\ps^{-1}$; medium green (medium grey)
  $22\ps^{-1}$--$250\ps^{-1}$; and
  dark green (dark grey) regions indicate points in paramter
  space where $\delmbs>250\ps^{-1}$. The soft sector is parameterised
  in the same manner as~\fig{bsm:cnt}.\label{bbb:2del}}
}
It has been known for some time~\cite{NH:th2,FOR:bdec}
that if flavour mixings appear in both the LL and RR sectors
then the contributions to $\delmbs$ can be especially large. Such a
situation is depicted in the left panel in~\fig{bbb:2del}.
Here we can see regions where $\delmbs$ exceeds values of
$250\ps^{-1}$ (a value unobservable at LHCb) and still satisfies the
current published D{\O} bound $\brbsm<5.0\times10^{-7}$.
Similar contributions arise
if flavour mixings appear in both the LR and RL sectors as illustrated
in the right panel in the figure. Here the constraints
and the general behaviour of the panel arise only when
one includes effects that appear beyond the leading order.
(One should note, however, that we do not include the constraint
supplied by $\bsg$ in both panels.) The large contributions
that appear in both panels only arise if flavour violation in the
LL or LR sectors occur together with flavour violation in the
RL or RR sectors.

\section{Limits on Single Sources of Flavour Violation}
\label{CUR:LimSing}

\FIGURE[t!]{
  \begin{tabular}{c c}
    \includegraphics[width=0.49\textwidth]{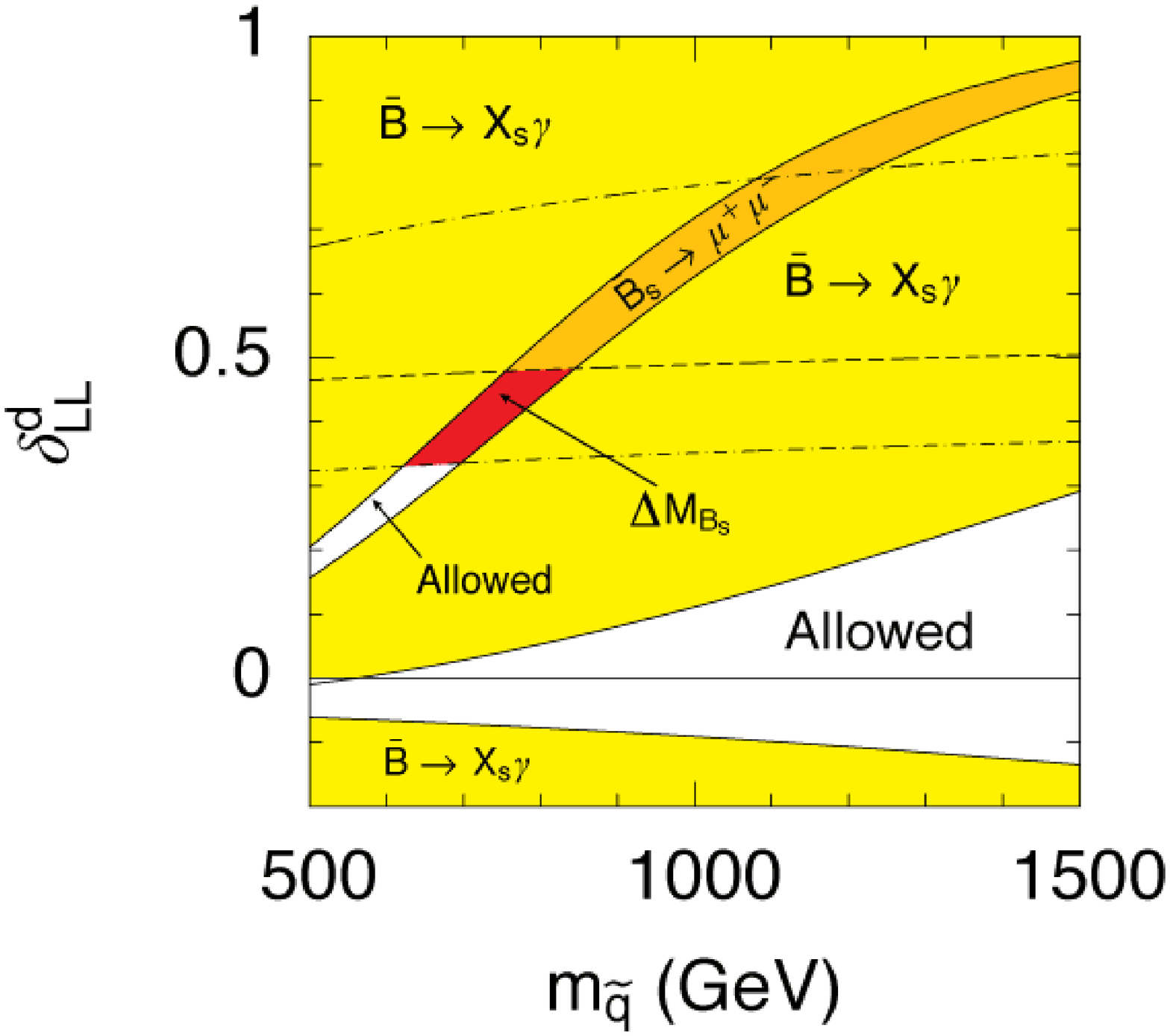}
    & \includegraphics[width=0.49\textwidth]{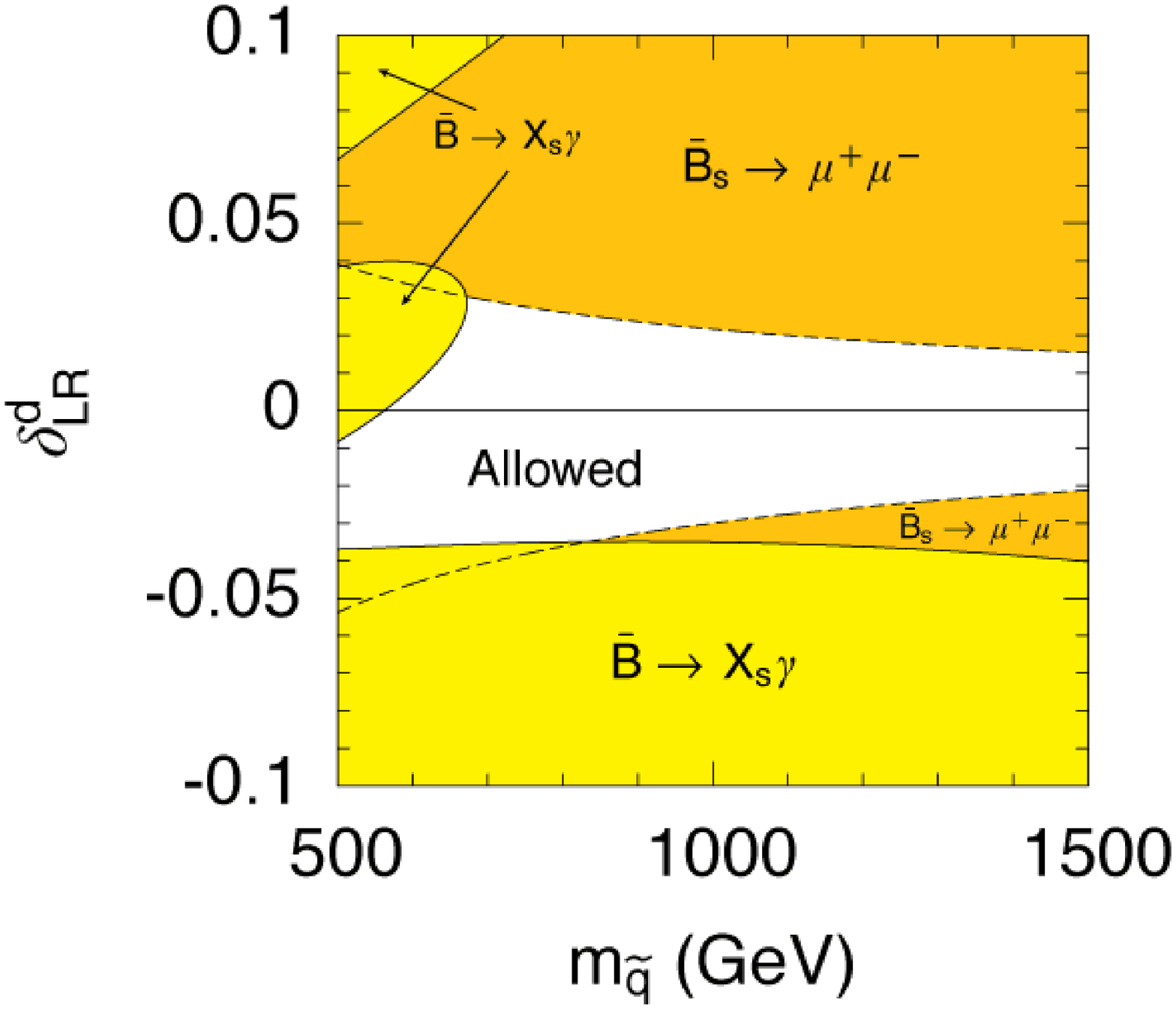}\\
    \includegraphics[width=0.49\textwidth]{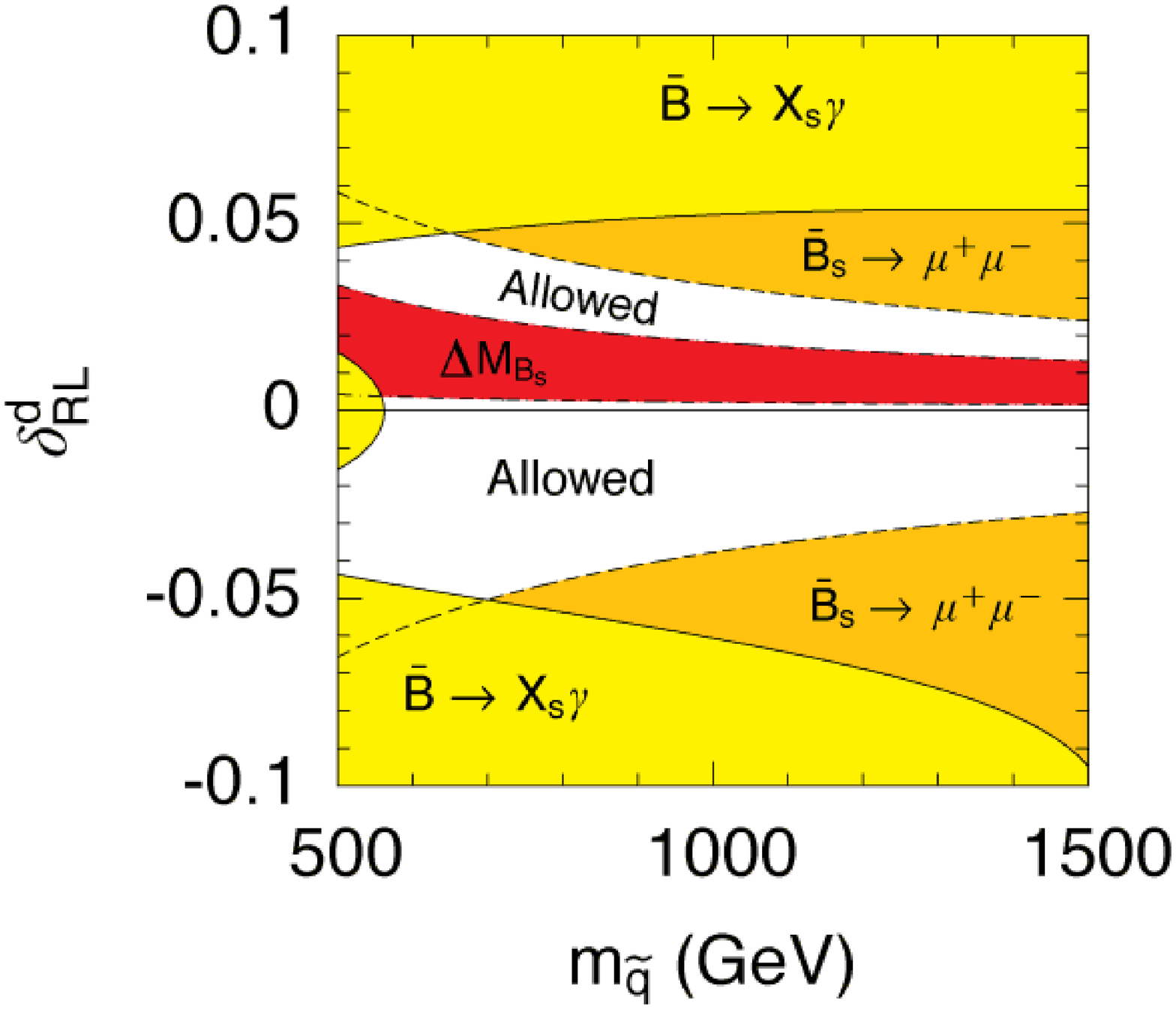}
    & \includegraphics[width=0.49\textwidth]{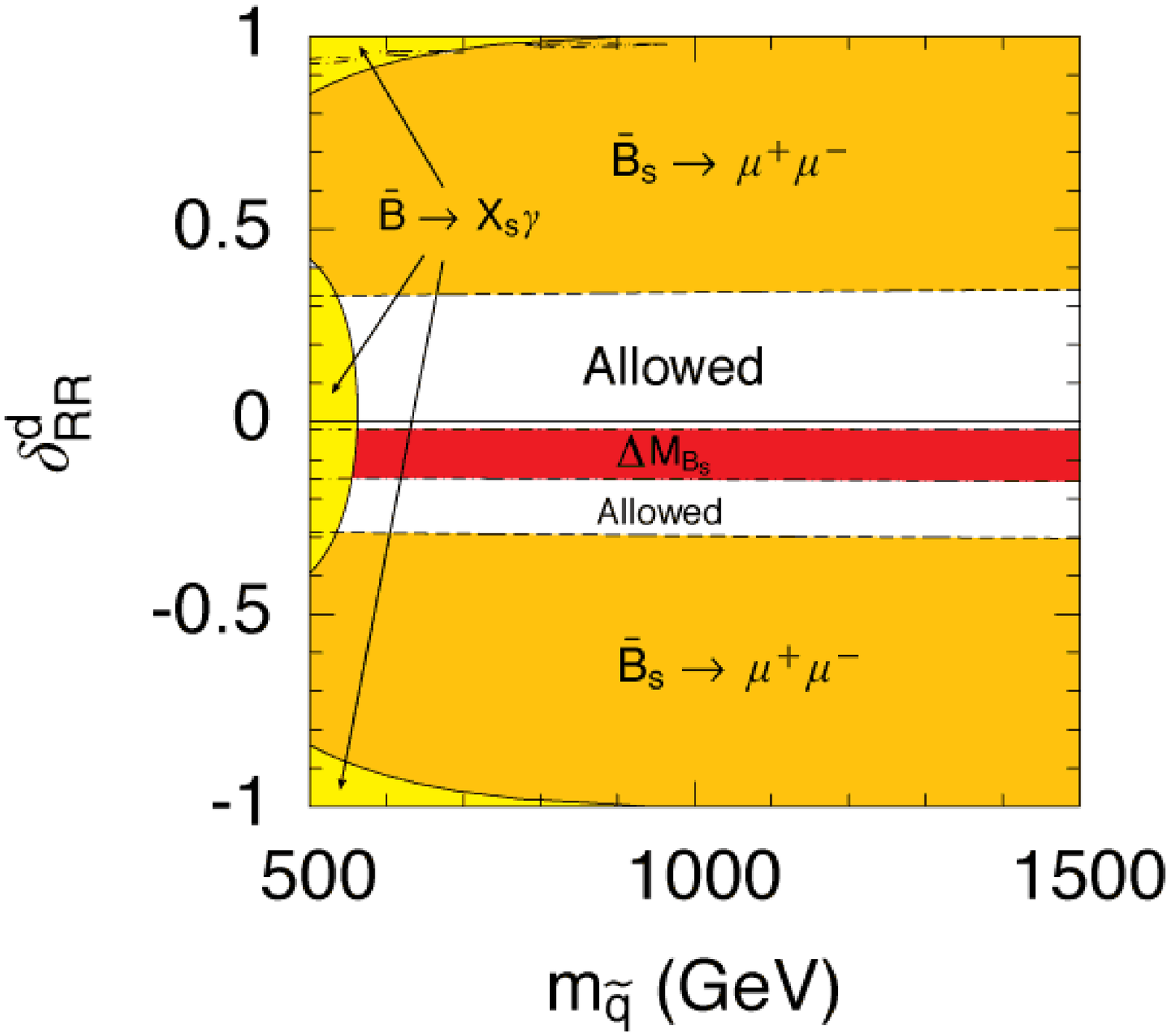}
  \end{tabular}
  \caption
  	{
  	Contour plots showing the current limits on the GFM parameters
	$\dxy$ for varying $\msq$ and fixed $m_A$. The soft sector
	is parameterised in a similar manner to~\fig{bsg:cnt}. Regions
	excluded by $\bsg$ are shaded in yellow (light grey), the
	subsequent regions that are excluded by $\bsm$ are shaded in
	orange (medium grey), finally the remaining regions that are 
	excluded by the limit on $\delmbs$ are shaded in red
	(dark grey)\label{Cur:cnt}.}
}
With the limits that can be provided by each process in mind, let us
now combine all three constraints discussed above. Such a situation
is shown in~\fig{Cur:cnt} where we vary $\msq$ along with a single
insertion. Unlike the figures shown in the previous three sections
we use white to depict allowed regions in parameter space (rather
than green/grey) and choose instead to shade the regions excluded
by each successive constraint we impose.

The top--left panel in the figure illustrates the allowed parameter
space for the insertion $\dll$. It is evident from the figure that
the constraints supplied by $\bsm$ and $\bbb$ mixing play
a useful r{\^o}le in constraining large values of $\dll$.
In particular, they tend to rule out the extreme regions of parameter
space, where the SUSY contributions to $\bsg$ have effectively
flipped the sign of the underlying amplitude relative to the
SM result. It is also apparent that, when one inspects the
contours that outline the regions allowed by these constraints,
both processes display a weak dependence on $\msq$. This
is due to the decoupling effect of the SUSY corrections
to the neutral Higgs vertex~\cite{NH:th,NH:th2,NH:rev}. These constraints will therefore remain even when $\msq$ approaches values for in
excess of $1\tev$ (provided that $m_A$ remains constant). 

The constraints on the insertion $\dlr$ are illustrated in the
top--right panel. In this plot we see the constraint supplied
by $\bsm$ once again tends to disfavour larger values of $\dlr$.
It is also apparent from the plot that the constraint seems
to become more severe with increasing $\msq$. This behaviour
can be readily understood by recalling that the insertion
$\dlr$ scales as $1/M_{SUSY}$. If one were to take into account
this behaviour somehow, by plotting the constraints on the off--diagonal
elements of the actual trilinear soft terms, for example, the correct
decoupling behaviour would be restored.

The bottom--left plot in~\fig{Cur:cnt} illustrates the constraints
on the insertion $\drl$. Here we see that the processes $\bsm$
and $\bbb$ remove additional ranges of $\drl$ for $\msq$
as small as $550\gev$ and determine the limits on the insertion
completely above $\msq\sim 700\gev$. Once again the limits imposed
by these constraints increase with $\msq$, however, in a similar
manner to the insertion $\dlr$ discussed above, it is apparent that,
once one takes into account the dependence of the insertion on 
the underlying SUSY mass scale, the correct decoupling
will be restored.

Finally, consider the bottom--right panel in the figure that
illustrates the bounds on the insertion $\drr$. Here we see
that, apart from $\msq\sim 500\gev$ -- where MFV is disfavoured
by the $\bsg$ constraint, the limits on the insertion are
determined completely by the limits on $\mathrm{BR}\left(\bsm\right)$
and $\delmbs$. In addition, in a similar manner to the insertion
$\dll$ the dependence of these two constraints on the underlying
SUSY mass scale is extremely weak. As such, the constraints
supplied by these two processes will continue to be useful, even if
the squarks and gluino effectively decouple.

\FIGURE[t!]{
  \begin{tabular}{c c}
    \includegraphics[width=0.49\textwidth]{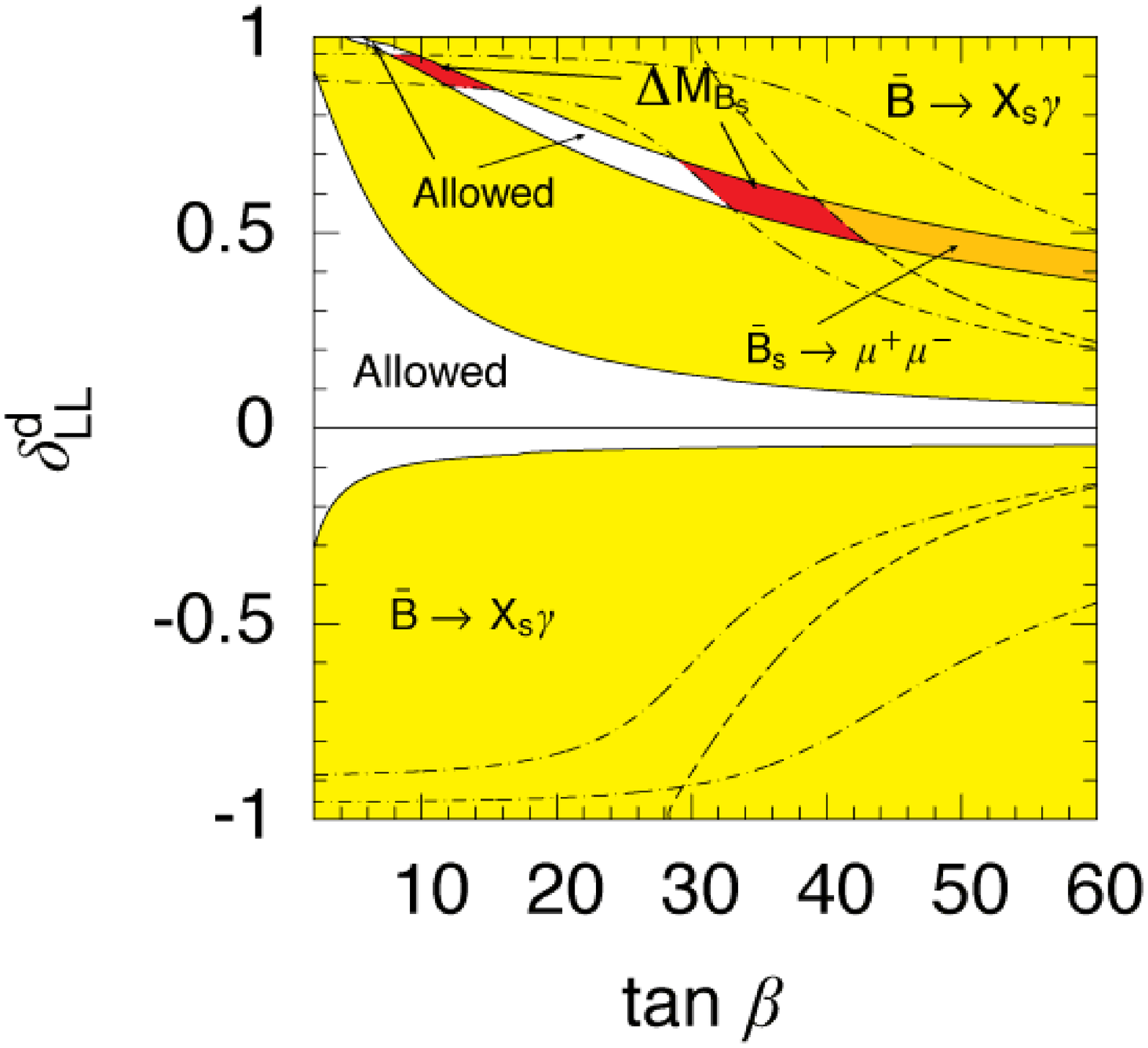}
    & \includegraphics[width=0.49\textwidth]{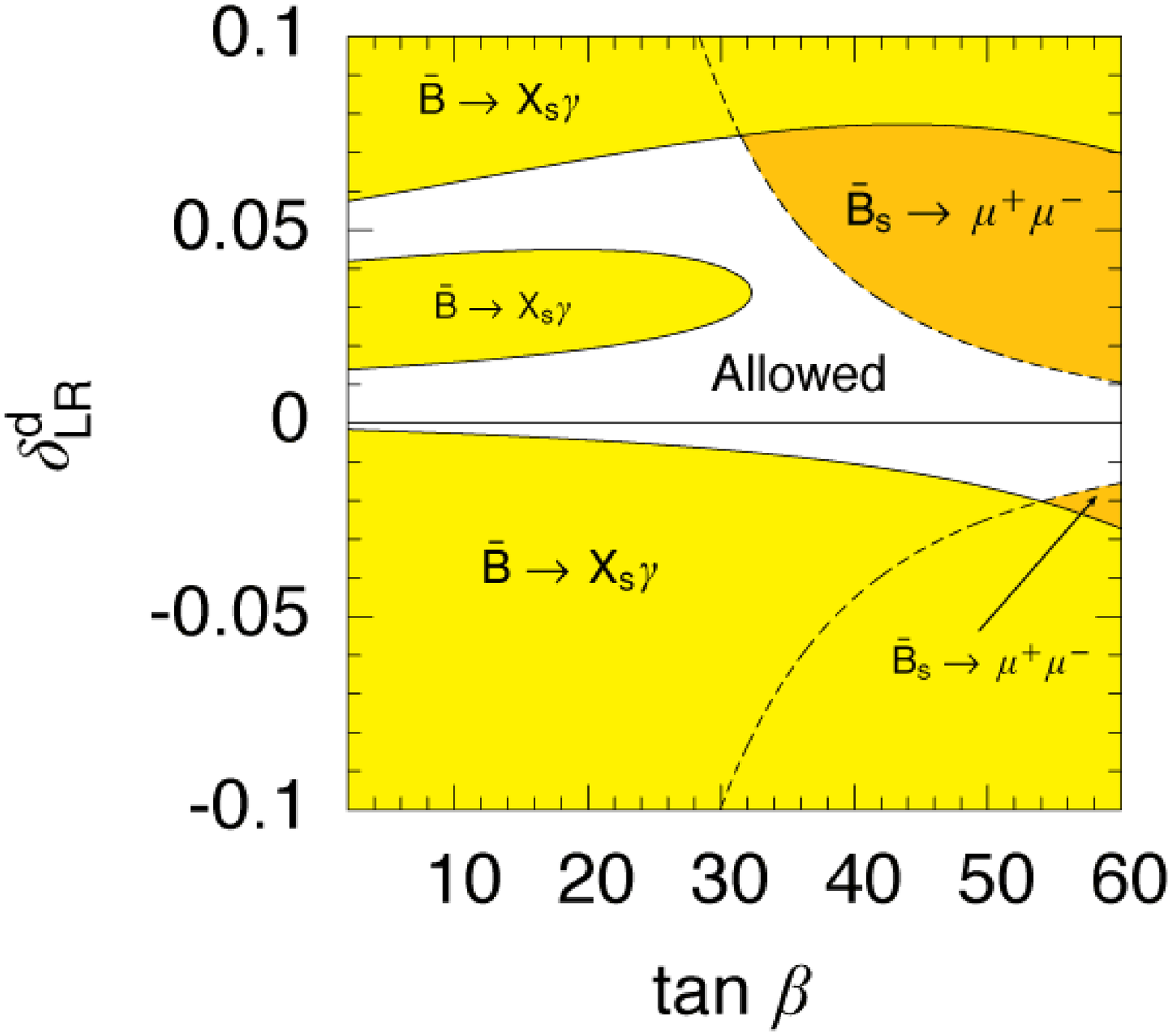}\\
    \includegraphics[width=0.49\textwidth]{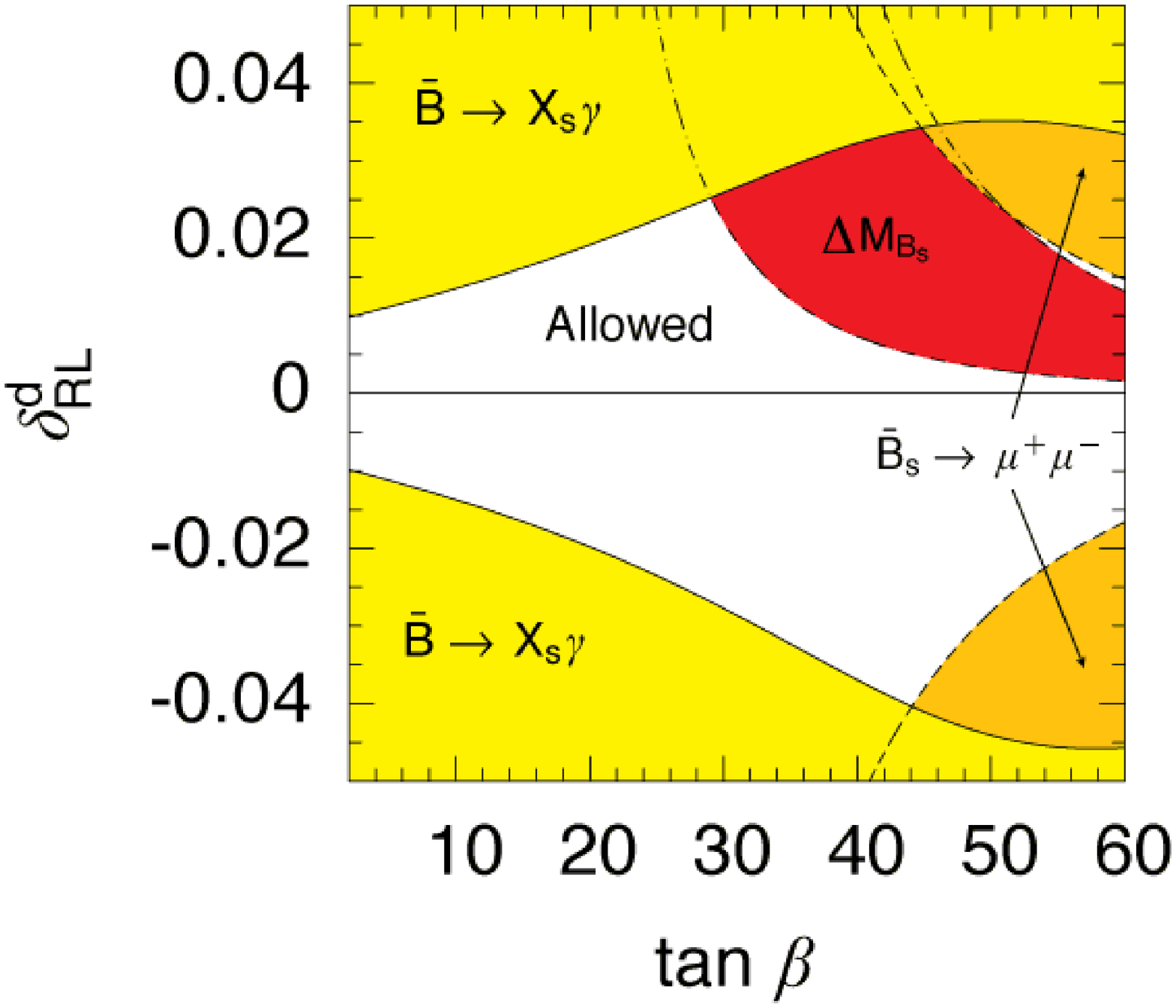}
    & \includegraphics[width=0.49\textwidth]{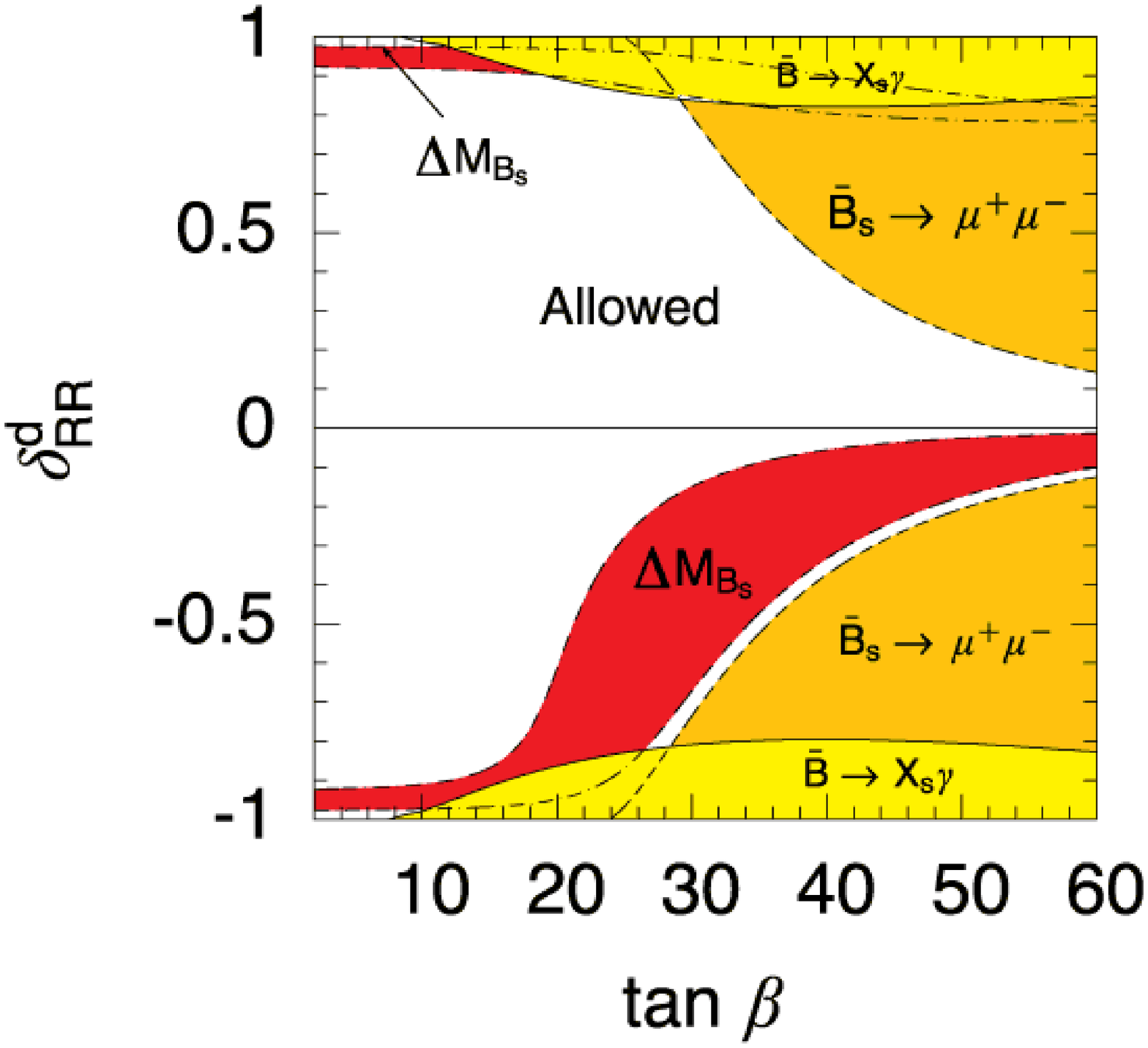}
  \end{tabular}
  \caption
  {Contour plots showing the limits on the GFM parameters
  $\dxy$ for varying $\tanb$. The soft sector is parameterised
  in a similar manner to~\fig{bsm:cnt} with $m_A=500\gev$. The
  excluded/allowed regions are shaded in a similar manner to~\fig{Cur:cnt}.
  \label{Cur:tanb}}
  }
The SUSY contributions to all three processes under investigation
in this paper are well known to be highly dependent on $\tanb$.
The corrections to $\mathrm{BR}\left(\bsm\right)$ and $\delmbs$
for instance scale as $\tan^6\beta$ and $\tan^4\beta$ while the
BLO focusing effect, discussed in~\cite{OR1:bsg,OR2:bsg,FOR:bdec} and
at the beginning of section~\ref{CUR:bsg}, is also strongly dependent
on $\tanb$. It is, therefore, natural to ask how the limits on the
various sources of flavour violation depend on $\tanb$. Such a
situation is shown in~\fig{Cur:tanb} where we illustrate the
allowed regions in the parameter space formed by varying $\tanb$
and one insertion at a time.

In the top--left panel the constraints on the insertion $\dll$
are shown. Here we see that, in contrast to the other three
panels in the figure, the constraints on the insertion increase
dramatically with $\tanb$. For example, at low $\tanb$,
the insertion is relatively unconstrained and values of up
to $\dll\sim 0.8$ are easily possible. This is because the
dominant LO chargino and gluino contributions to the decay
that arise from LL insertions are both proportional to
$\tanb$. As $\tanb$ increases the constraint
imposed by $\bsg$ therefore becomes much more stringent
and separates into two distinct branches. The upper
branch here is associated with regions where the amplitude
associated with the decay has effectively changed sign.
Once we include the constraints supplied by $\bsm$ and $\bbb$
mixing, however, these extreme regions are typically ruled
out for $\tanb>30$ (and also, to some extent, for $\tanb<15$).

The limits on the insertion $\dlr$ are shown in the top--right
panel. Here we see the focusing effect rather clearly:
the regions of parameter space excluded by the $\bsg$ constraint
decrease dramatically as $\tanb$ increases, in contrast to a LO
analysis where, essentially, the 
bounds on $\dlr$ are relatively independent of
$\tanb$. As $\tanb$ increases, however, the constraint
supplied by the decay $\bsm$ also gains prominence and
becomes important in determining the limits on the insertion
above $\tanb\sim30$. Indeed, above $\tanb\sim 55$,
the constraints on $\dlr$ are determined entirely
by the published D{\O} bound $\brbsm<5.0\times10^{-7}$.

The lower--left panel in the figure illustrates the bounds on
the flavour violation in the RL sector. Once again, the
bounds on the insertion supplied by the $\bsg$ constraint
decrease steadily as $\tanb$ increases. This is due
to the focusing effect discussed at the beginning
of section~\ref{CUR:bsg}. It is also apparent from the figure 
that the constraints supplied by $\bsm$ and $\bbb$ mixing
play large r{\^o}les when determining the limits
on $\drl$ at large $\tanb$. The limits imposed by the
current lower bound on $\delmbs$, for instance, start
eliminating large regions of parameter space for positive
$\drl$ for $\tanb$ as low as $30$. Increasing $\tanb$
only serves to increase the effectiveness of the
constraints supplied by the decay $\bsm$ and $\bbb$ mixing,
and, for example, above $\tanb\sim 45$ the constraint
provided by $\bsg$ is entirely supplanted by those
attributable to $\bsm$ and $\bbb$ mixing. 

Finally, in the bottom--right panel we show the limits on the
insertion $\drr$. From the plot it is evident that the
bounds imposed by $\bsg$ are fairly weak. This is principally
due to two reasons: the first is that the insertion
$\drr$ only contributes (in a significant manner) to the
primed Wilson coefficients and, therefore, cannot interfere
directly with the SM contribution; the second is that the 
BLO focusing effect significantly reduces the
$\tanb$ enhanced LO contribution that arises from gluino
exchange. It is therefore apparent from the figure that
the constraints supplied by both $\bsm$ and $\bbb$ 
can become important even for $\tanb$ as low as $20$ 
and for $\tanb>30$ the constraints on the insertion
are determined entirely by these two processes.

\begin{table}[t!]
\small{
\begin{center}
\begin{tabular}{c| l l l}
\hline
\hline
$\dll (10^{-2})$ & $\tanb=10$ & $\tanb=40$ & $\tanb=50$\\
\hline
\hline
A & $(-2.2,11.0),\,(47,56)$ & $(-4,11)$ & $[-1.2,-0.2)$\\
B & $(-10,40)$ & $(-7,10)$ & $(-7.4,8.0)$\\
C & $(-5,8.6),\,(45.6,54.8)$ & $(-4.4,0.2),\,(14.4,18.8)$ & $(-4.6,-0.7),\,(12.0,15.8)$\\
D & $(-17.4,37.0)$ & $(-9.2,9.6)$ & $(-8.8,7.4)$\\
\hline
\hline
$\dlr (10^{-3})$ & $\tanb=10$ & $\tanb=40$ & $\tanb=50$\\
\hline
\hline
A & $(-1.2,7.6),\,(22,31)$ & $(-12,2)$ & $[-9.2,-1.6)$\\
B & $(-3,15),\,(46,64)$ & $(-12,18)$ & $[-14.8,9.2]$\\
C & $(-3.0,5.4),\,(21.0,29.6)$ & $(-13.4,0.2),\,(26.4,33.6]$ & $(-19.2,-3.6)$\\
D & $(-5.00,12.8),\,(45.0,63.0)$ & $(-14.8,19.6),\,(55.2,78.8]$ & $(-20.2,24.4)$\\
\hline
\hline
$\drl (10^{-3})$ & $\tanb=10$ & $\tanb=40$ & $\tanb=50$\\
\hline
\hline
A & $(-7.0,7.0)$ & $[-12,0.5\},\,\{6,10]$ & Excl.\\
B & $(-16,16)$ & $[-30,2\},\,\{14,26]$ & $[-16.0,0.6\},\,\{6.6,14.0]$\\
C & $(-9.8,9.8)$ & $(-32.8,3.4\},\,\{26.2,32.0)$ & $[-28.0,-15.0),\,(13.8,24.8]$\\
D & $(-19.4,19.4)$ & $(-52.8,7.6\}$ & $[-64.4,3.4\},\,\{27.6,57.2]$\\
\hline
\hline
$\drr (10^{-2})$ & $\tanb=10$ & $\tanb=40$ & $\tanb=50$\\
\hline
\hline
A & $(-24.6,26.8)$ & $[-5,-2\},\,\{-0.4,6]$ & Excl.\\
B & $(-80,84)$ & $[-23.2,-11.4\},\,\{-1.6,26.4]$ & $[-13.0,-6.0\},\,\{-0.6,14.8]$\\
C & $(-36.4,38.6)$ & $[-22.6,-11.6\},\,\{-1.6,25.8]$ & $[-12.6,-6.1\},\,(9.4,14.6]$\\
D & $\{91.2,92.0\}$ & $(-71.4,-39.2\},\,\{-6.4,74.2)$ & $[-50.8,-21.8\},\,\{-3.1,56.8]$\\
\hline
\hline
\end{tabular}
\end{center}
}
\caption
{Table of limits on all four insertions for varying $\tanb$.
The soft sector is parameterised as follows: in all four scenarios
$\msq=\mgl=\mu$ and $A_u=-\msq$; in scenarios A and C $\msq=500\gev$;
in scenarios B and D $\msq=1\tev$. The pseudoscalar mass $m_A$
is given by $m_A=\msq/2$ in scenarios A and B, and by $m_A=\msq$
for scenarios C and D. The source of each bound is indicated
by varying styles of bracket, conventional round brackets indicate $\bsg$,
square brackets $\bsm$ and braces indicate $\bbb$ mixing.\label{Cur:tab}}
\end{table}
To illustrate the usefulness of combining the constraints
supplied by $\bsg$, $\bsm$ and $\bbb$ mixing we provide a list
of the bounds on each insertion in table~\ref{Cur:tab}. The
first column in the table illustrates the bounds on the 
insertions for $\tanb=10$. It is evident here that,
for this moderate value of $\tanb$, the constraints
on all four insertions are determined entirely by the decay
$\bsg$ (except for one case for $\drr$). This is, of course, to be
expected due to the large dependence the neutral Higgs penguin
contributions to $\bsm$ and $\delmbs$ have on $\tanb$.
Moving on to the next two columns which display the limits
on the insertions for much larger values of $\tanb$ it can
be seen that both $\bsm$ and $\bbb$ mixing play a much more
useful r{\^o}le. For example, for LL and LR insertions
the constraints either remove the second set of allowed values that
were present at $\tanb=10$ or reduce them somewhat, even for relatively
large $m_A$. For RL and RR insertions the constraints are determined
almost entirely (with one or two exceptions) by the current limits
on $\mathrm{BR}\left(\bsm\right)$ and $\delmbs$. The allowed values
of these insertions also take a different form compared
with those obtained for small $\tanb$. For low $\tanb$ the
allowed regions are usually given by (roughly) symmetric regions centred
on the MFV limit (\ie~$\dxy=0$). However, at large $\tanb$
the allowed regions appear to split into two separate branches.
These branches that appear at negative $\drr$ and positive $\drl$
are, essentially, regions of parameter space where the SUSY
contributions have flipped the sign of the underlying amplitude,
relevant to $\bbb$ mixing, relative to the SM result.

\FIGURE[t!]{
  \begin{tabular}{c c}
    \includegraphics[width=0.49\textwidth]{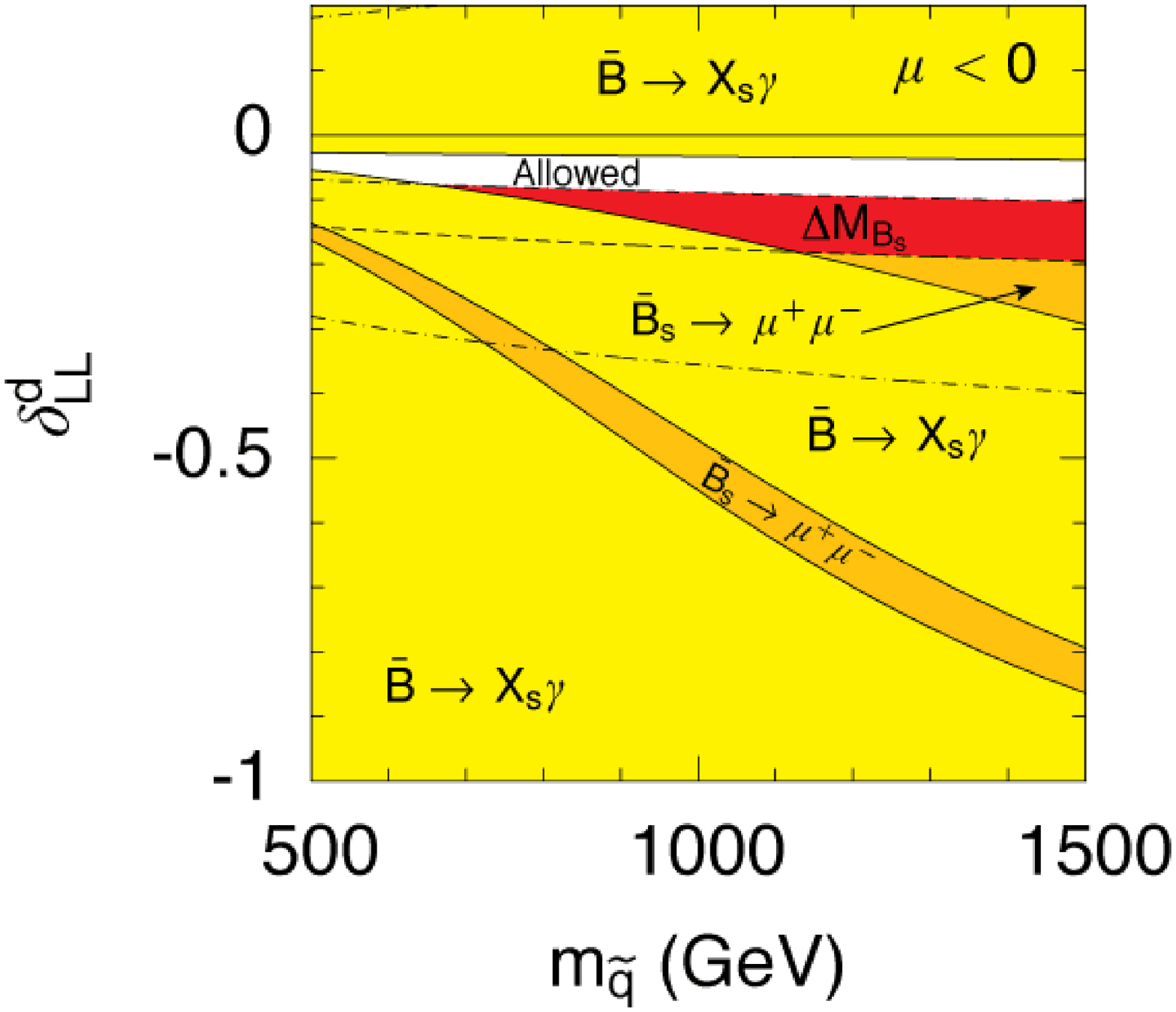}
    & \includegraphics[width=0.49\textwidth]{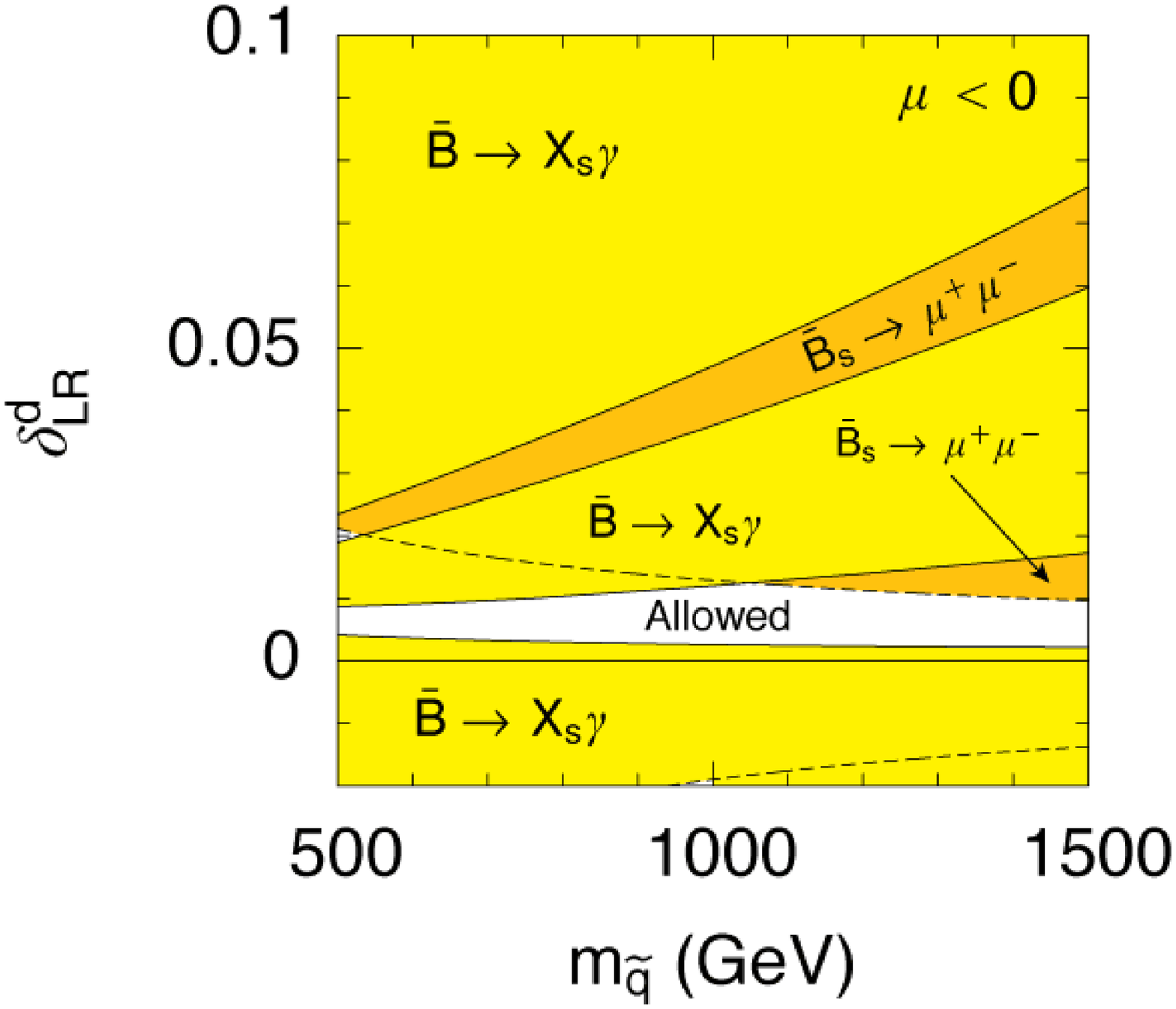}
  \end{tabular}
  \caption
  	{
  	Contour plots showing the current limits on the GFM parameters
	$\dll$ and $\dlr$ for varying $\msq$ and $\mu<0$. The soft sector
	is parameterised in a similar manner to~\fig{bsg:cnt} while the
	shading of the figure is that same as~\fig{Cur:cnt}.\label{Cur:negmu}}
}
Let us now consider the case $\mu<0$. As discussed earlier
in this paper, for $\mu<0$, the chargino and charged Higgs
contributions to $\bsg$ interfere constructively with the SM
result and one is often forced to adopt a very heavy SUSY
mass spectrum ($M_{SUSY}\gg 1\tev$) to ensure that the contributions
to $\brbsg$ are not too large. In GFM it is possible to improve the
situation by considering contributions from either
LL or LR insertions that interfere destructively with the chargino
and charged Higgs corrections to the decay. This problem is made
somewhat easier when one considers BLO effects as, instead of reducing
the GFM contribution, as they do for positive $\mu$, they act to 
increase it (\ie~``anti--focusing''). As such it is possible that
relatively small deviations from MFV will satisfy the $\bsg$ constraint.
As changing the sign of $\mu$ also changes the sign of the term
$\epsilon_s$ that appears in the various BLO corrections presented
in~\cite{FOR:bdec}, it is apparent that the contributions to both
$\bsm$ and $\bbb$ mixing will increase as well. It is therefore to be
expected that the constraints imposed by these two processes will
play an even greater r{\^o}le compared to the positive $\mu$ case.
Such a situation is shown in~\fig{Cur:negmu} where we present the allowed
parameter space in the $\dll$--$\msq$ and $\dlr$--$\msq$ planes for
negative $\mu$. Each plot displays the familiar structure of two
branches that are compatible with the $\bsg$ constraint encountered
earlier in~\fig{bsg:cnt}. In a similar manner to the plots found
in~\fig{Cur:cnt} the constraints supplied by the limits on $\brbsm$
and $\delmbs$ tend to rule out the more extreme regions of parameter
space. For example in each figure the constraint supplied $\bsm$
excludes the extreme branch of the region permitted by $\bsg$. In
the left panel of~\fig{Cur:negmu} the constraint supplied
by $\bbb$ mixing also plays a large role and rules out over
half of the allowed parameter space in the upper branch in the
figure. It is obvious from both figures however that regions
consistent with all three constraints are still possible and
$\mu<0$ can still be a viable possibility at least when one considers
minor departures from MFV.

\FIGURE[t!]{
  \begin{tabular}{c c}
    \includegraphics[width=0.49\textwidth]{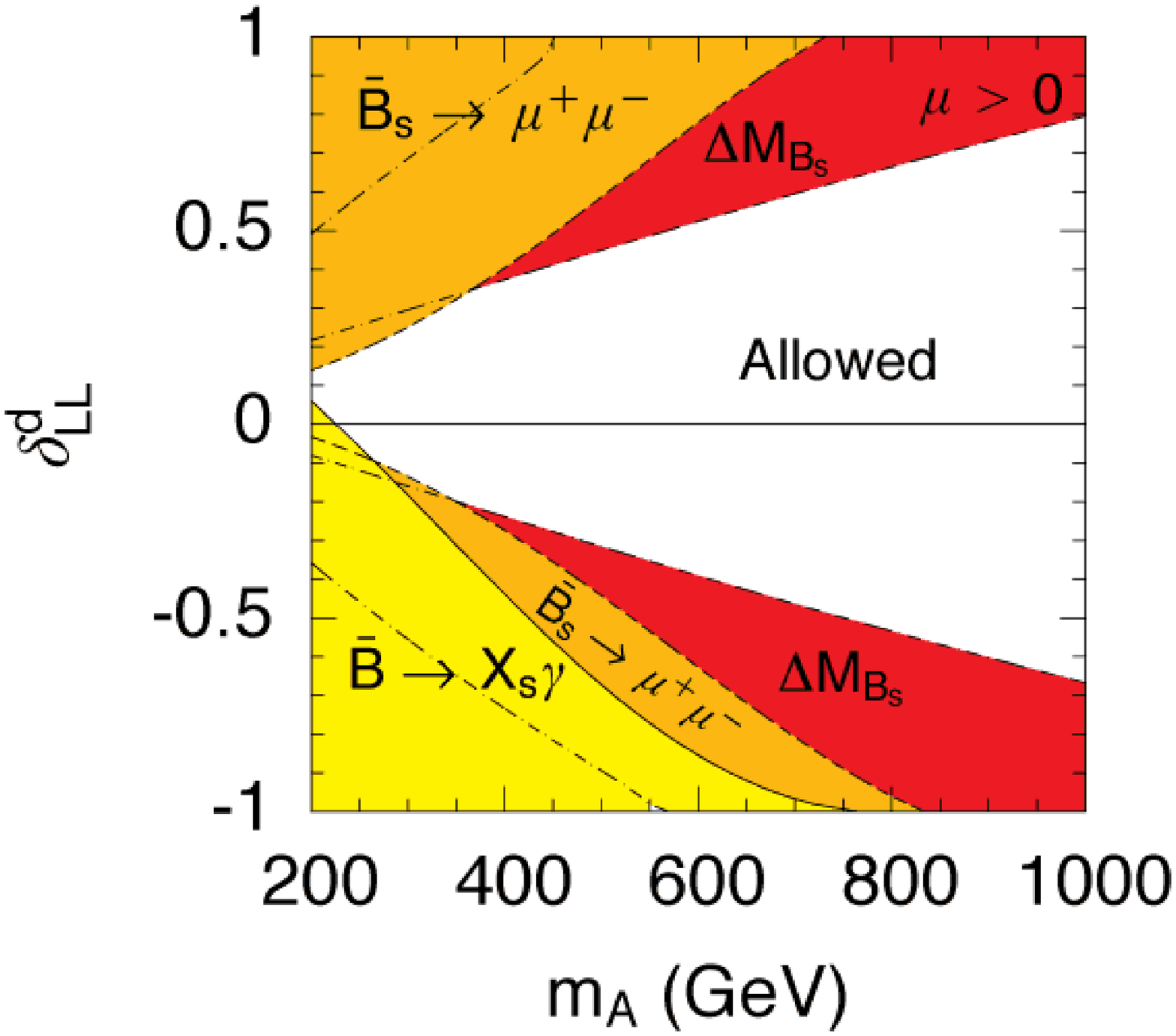}
    & \includegraphics[width=0.49\textwidth]{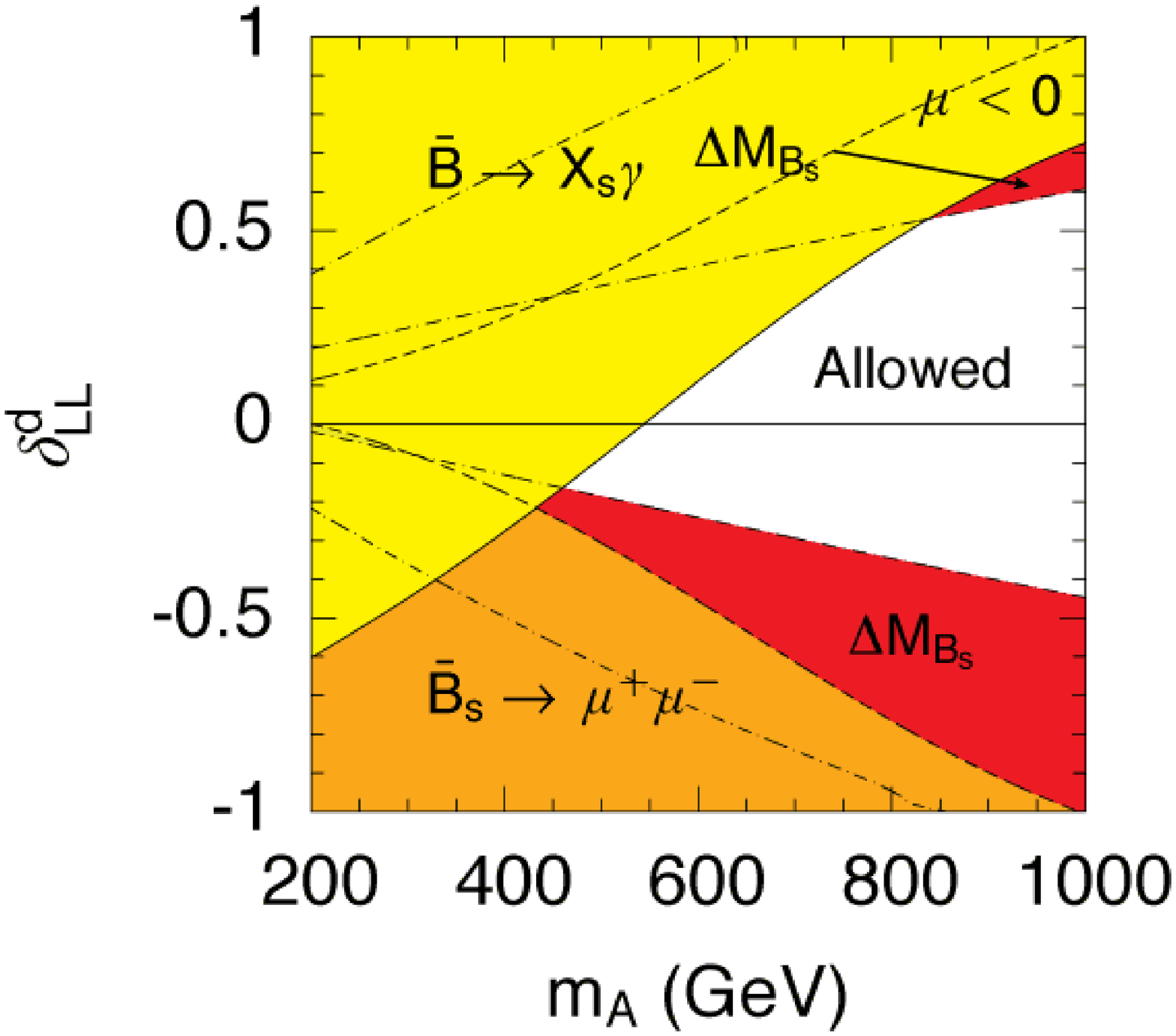}
  \end{tabular}
  \caption
  	{
  	Contour plots depicting the limits on the insertion $\dll$ in
	the decoupled limit $\msq\gg m_A$. The soft sector is parameterised
	as follows $\msq=\sqrt{2}\mu=\mgl/\sqrt{2}=-A_u=10\tev$ and $\tanb=40$.
	In the left panel $\mu>0$, while in the right panel $\mu<0$.
	\label{Cur:dc}}
}
Finally, let us consider the decoupled limit $\msq,\mgl\gg m_A$. Such
a situation is illustrated in~\fig{Cur:dc} where we depict the constraints
that can be placed on the insertion $\dll$. As is evident from the two
panels all three processes continue to play a r{\^o}le in constraining
the LL insertions. It should be noted that the constraints arising
from $\bsg$ in particular, are a purely BLO effect. The constraints
arise from the threshold corrections to the charged Higgs vertex 
and can act to either increase or decrease the contribution to 
$\bsg$ arising from charged Higgs exchange relative to a purely 
LO calculation. This is in contrast to an MFV calculation where the
BLO corrections act only to decrease the charged Higgs contribution
for $\mu>0$ and increase it for $\mu<0$. It is also evident from both
panels that the constraints provided by $\bsm$ and $\delmbs$, which
arise from the threshold corrections to the neutral Higgs vertex,
continue to play an important r{\^o}le eliminating roughly symmetric
regions of parameter space in each panel. Finally, let us note that
in each panel the lower bound on $m_A$ in the MFV limit
is determined solely by $\bsg$, however, once one proceeds beyond
this approximation the lower limit on $m_A$ can be dictated
by either the $\bsm$ or $\bbb$ constraints (especially
when $\mu>0$, as in the left panel).

It is natural to ask how the other insertions are constrained in this
extreme region of parameter space. As the threshold corrections
to the charged Higgs vertex, arising from RR insertions, only effect
the primed Wilson coefficients~\cite{FOR:bdec} it is apparent that
the constraint supplied by $\bsg$ will have a relatively minor
dependence on $\drr$. As such
the constraints on the insertion are usually determined by $\bsm$
and $\bbb$ mixing, and eliminate regions of parameter space
similar to those found in~\figs{bsm:cnt}{bbb:cnt}. Turning to flavour 
violation in the LR and RL sectors, as the original definitions
for these scale as $1/M_{SUSY}$, it is generally more useful
to constrain the actual elements of the trilinear soft terms
themselves, or at least a dimensionless quantity that does
not feature a Higgs VEV or quark mass. Provided one takes this
into account it is apparent from the decoupling behaviour exhibited
by the $\bsm$ and $\bbb$ constraints in~\fig{Cur:cnt} that both processes
will continue to be useful when constraining flavour violation
in the LR and RL sectors, even for $\msq=10\tev$. Turning to the
$\bsg$ constraint, neither insertion induces large threshold
corrections to the charged Higgs vertex and as such the dependence
of the $\bsg$ constraint on these two insertions would be expected
to be relatively small when compared to the large dependence on $\dll$
exhibited in~\fig{Cur:dc}.

\section{Limits on Multiple Sources of Flavour Violation}
\label{CUR:Multi}

\FIGURE[p!]{
  \begin{tabular}{c c}
    \includegraphics[width=0.48\textwidth]{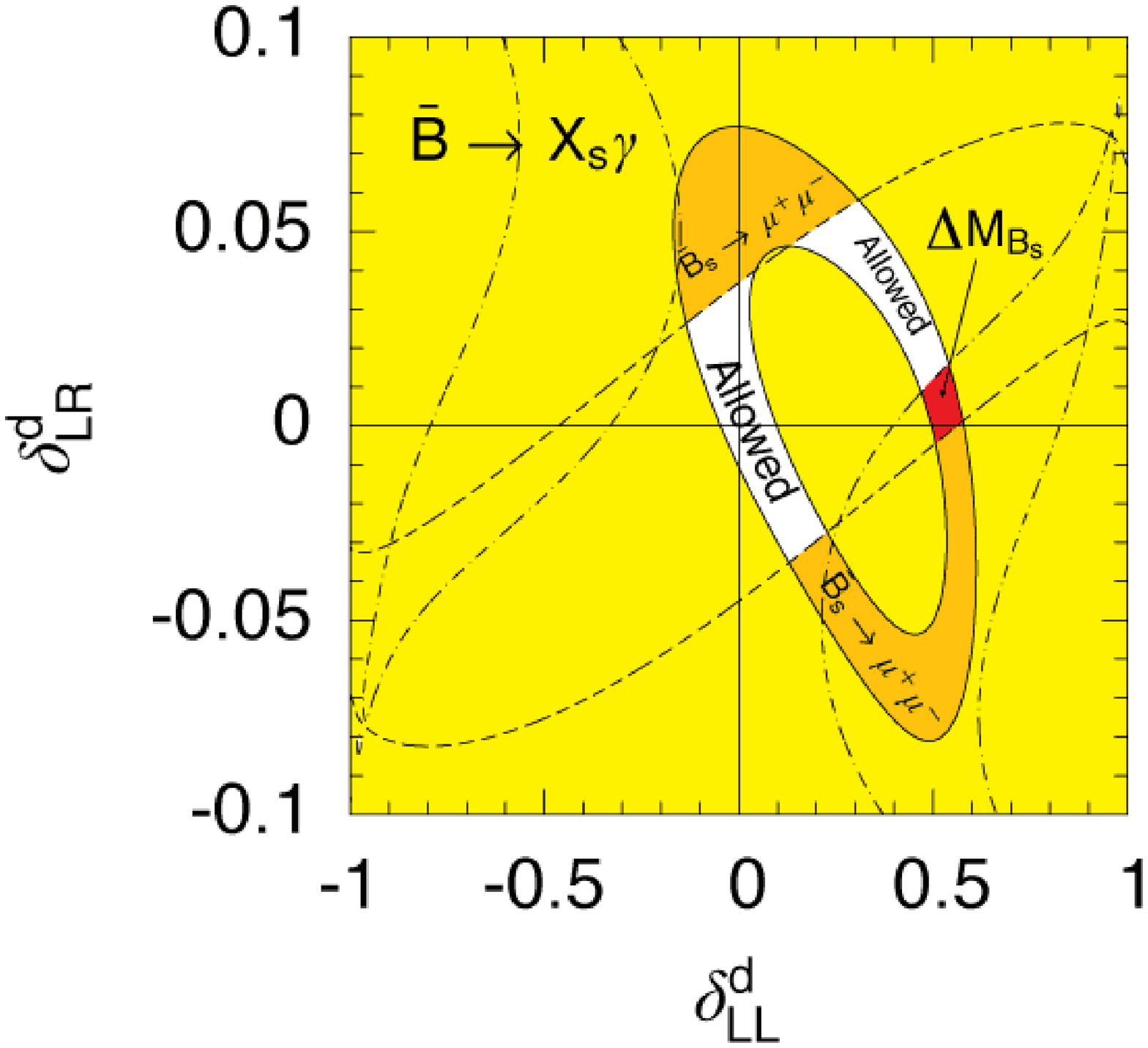}
    & \includegraphics[width=0.48\textwidth]{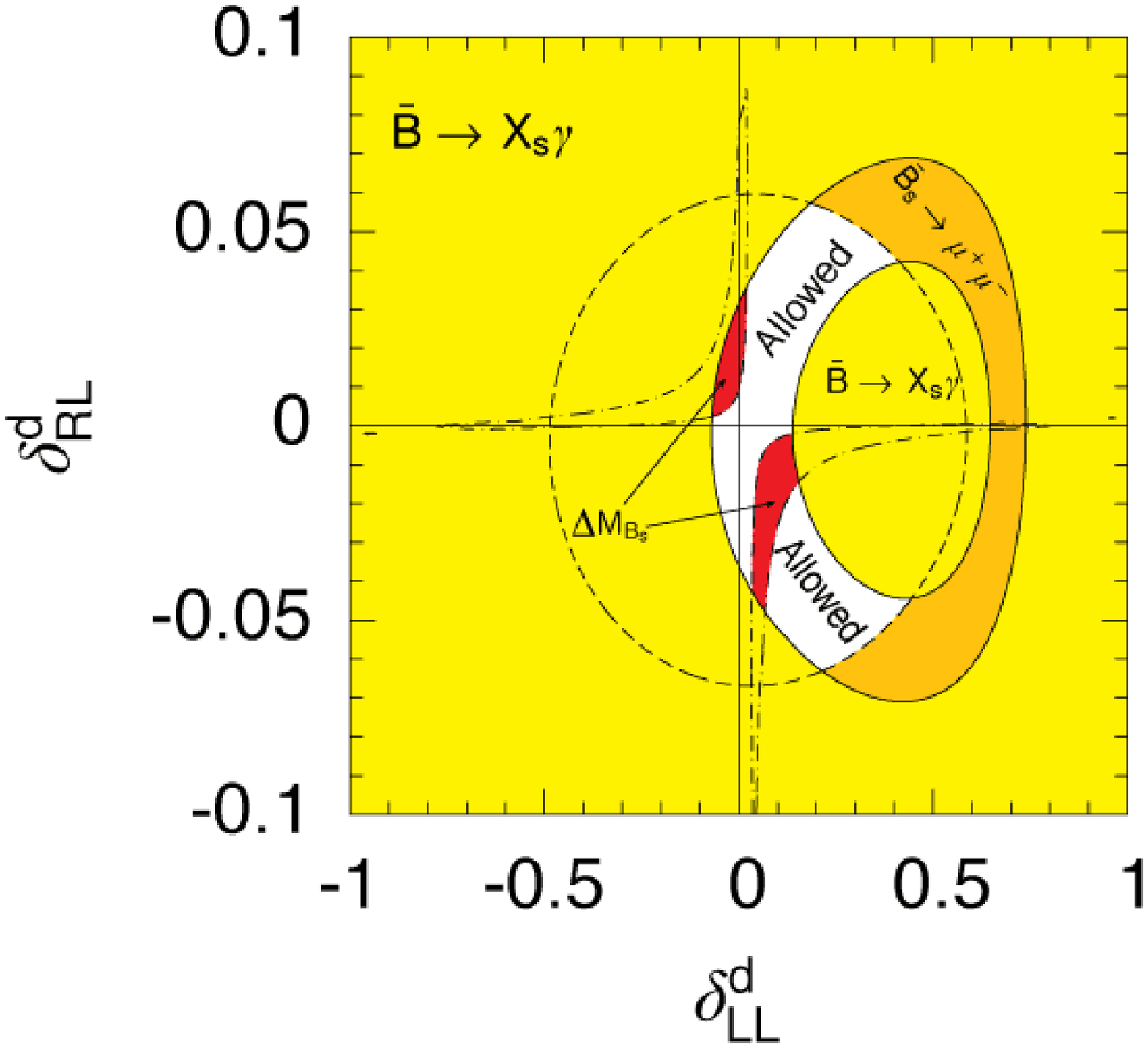}\\
    \includegraphics[width=0.48\textwidth]{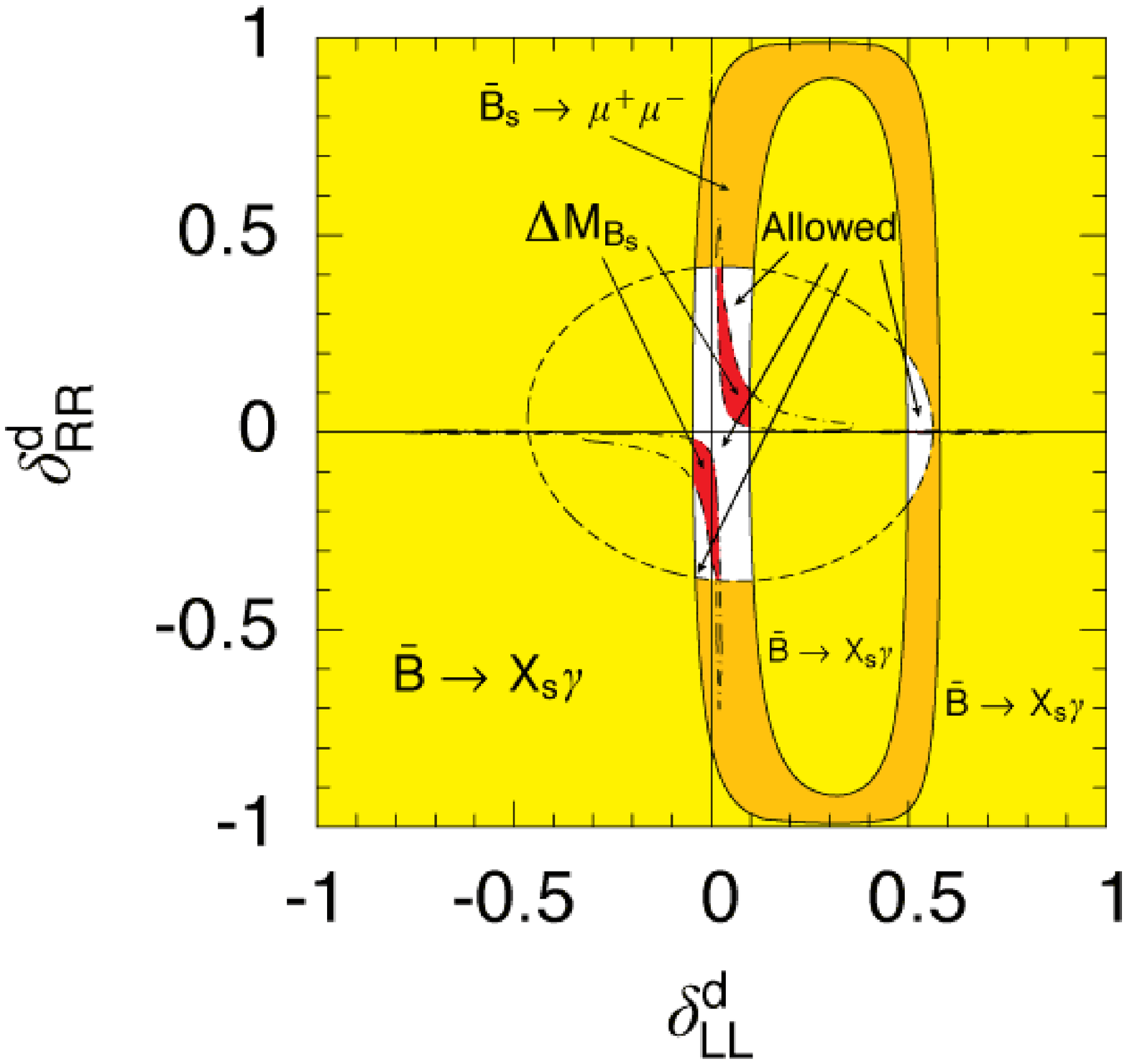}
    & \includegraphics[width=0.46\textwidth]{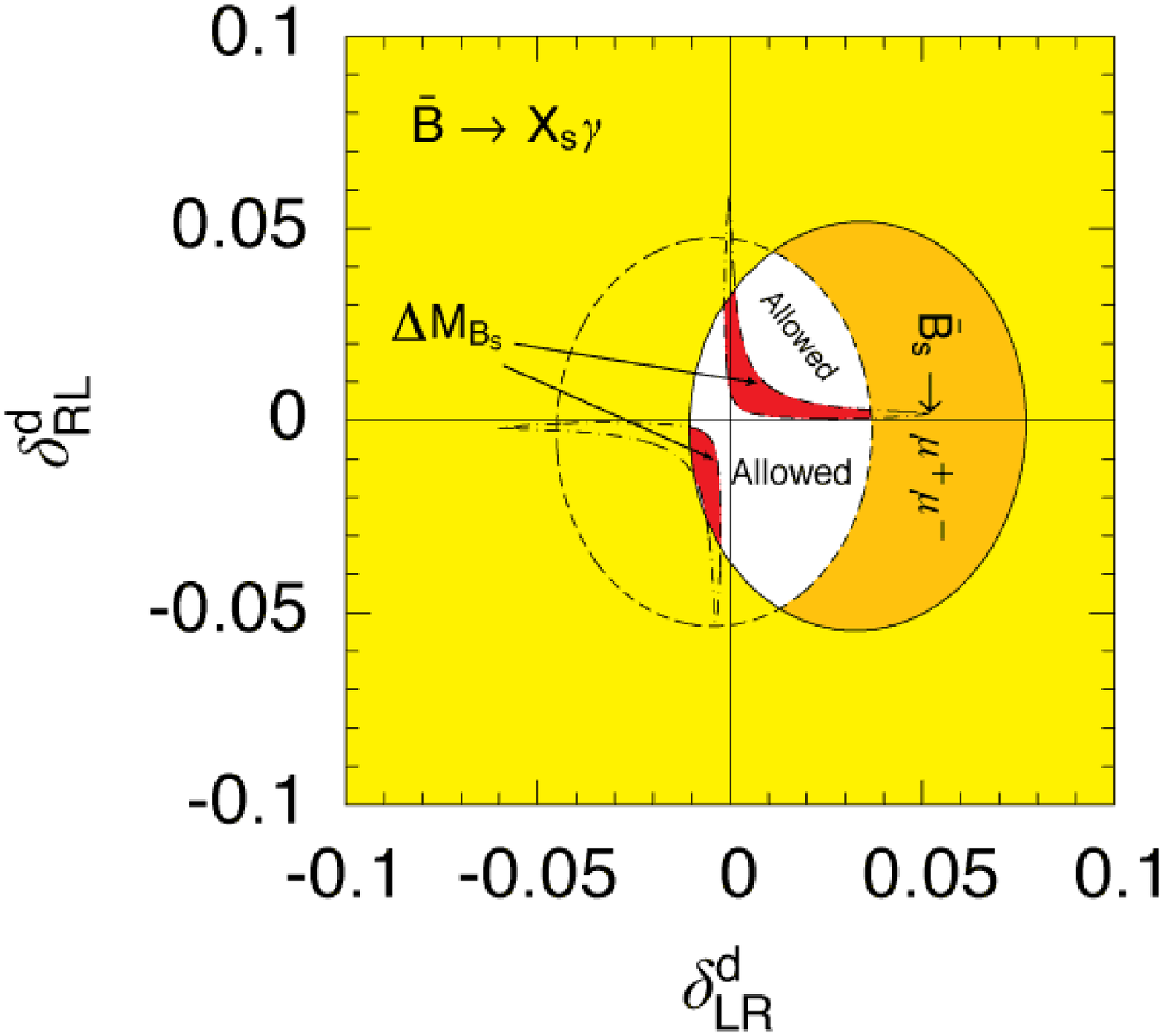}\\
    \includegraphics[width=0.48\textwidth]{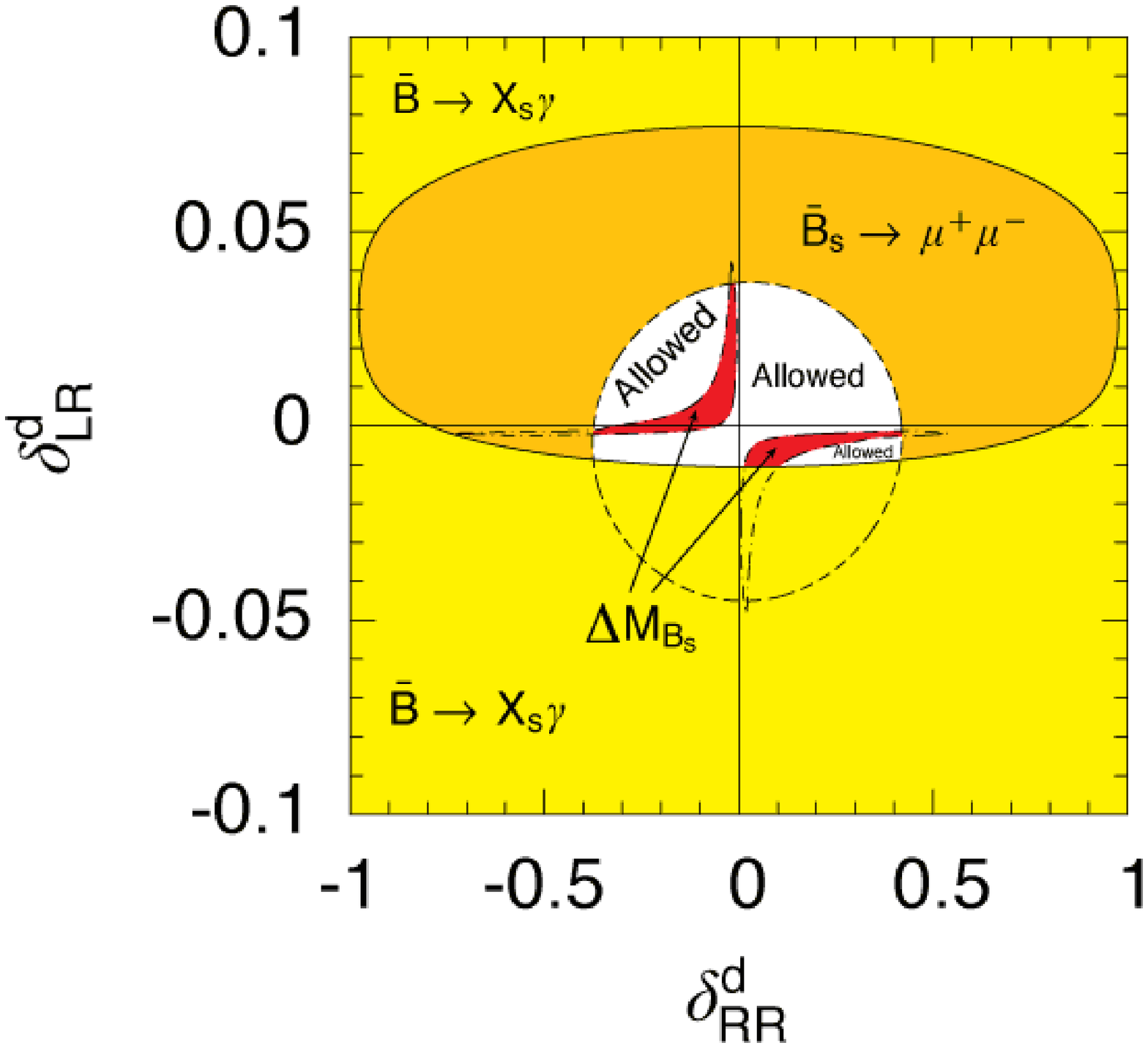}
    & \includegraphics[width=0.48\textwidth]{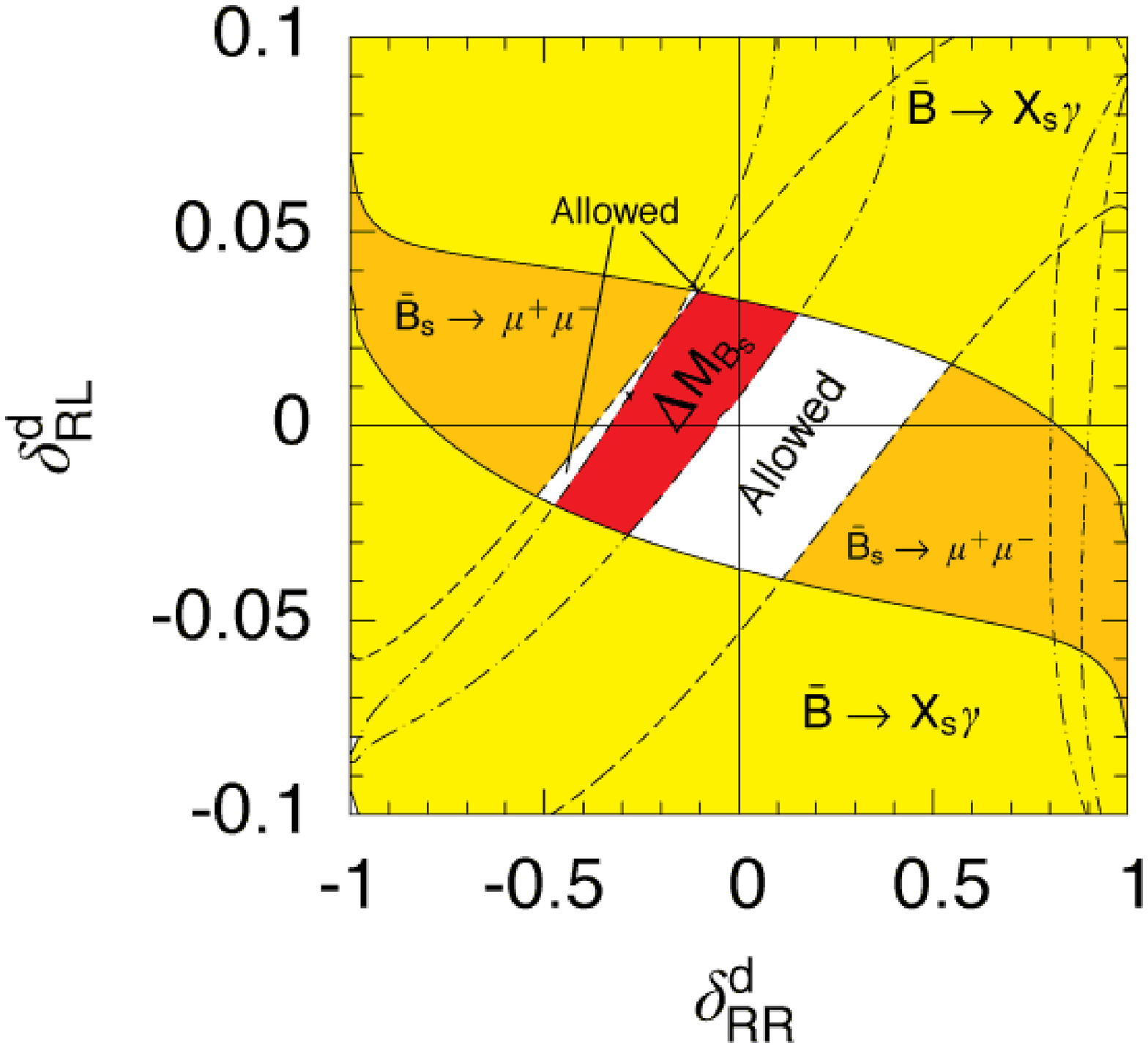}
  \end{tabular}
  \caption{Shading the same as \fig{Cur:cnt}, the soft sector
  is parameterised in a similar manner to~\fig{bsm:cnt} with
  $m_A=500\gev$.\label{Cur:mult}}
}
Up until now, for the sake of simplicity,
we have been mainly concerned with one source
of flavour violation being varied at a time. However, it is more
natural to expect more than one insertion to be simultaneously
non--zero. This may lead to a more varied picture.
For example, as discussed at the end of section~\ref{CUR:bbb},
large contributions to $\delmbs$ are possible if two
insertions are both non--zero. Coupled with the possibility
that interference between the various contributions can also
play a r\^ole for $\bsg$ and $\bsm$, it is interesting to consider
the limits attainable if one decides to vary two sources of flavour
violation in the squark sector at once. Such a situation is
shown in~\fig{Cur:mult}. 

First let us consider the regions ruled out by the $\bsg$
constraint (the yellow regions). The regions permitted by the constraint
are typically circular in nature with the exception of the
$\drl$--$\drr$
plane (the bottom--right panel). The reason
for the deviation is due to the fact that neither insertion
contributes significantly to the unprimed Wilson coefficients
$C_7$ and $C_8$ and the overall effect of the insertions
is to increase the branching ratio for the decay. For the
remaining combinations at least one of the insertions contributes
to the unprimed sector and can therefore lead to a decrease as well
as increase in BR$(\bsg)$. Let us briefly also
mention why some of the circular regions are filled (mainly
those featuring the insertion $\dlr$) whilst others are unfilled
(those that feature the insertion $\dll$). The reason here is due
to the focusing effect discussed 
at the beginning of section~\ref{CUR:bsg}. In a similar manner to the
top--right panel in~\fig{bsg:cnt}, the focusing effect raises the
minimum value for $\brbsg$ in the circular regions towards the SM
value and above the lower limit we impose
(a more stringent lower limit
would lead to ``doughnut'' shaped regions like the remaining panels).
As the contributions due to LL insertions are less affected by
BLO corrections (compared with the other three insertions)
the allowed regions in the plots featuring
this insertion tend to retain a similar form to their LO counterparts.

All six panels in the figure easily illustrate the usefulness
of the current bound in $\bsm$ in constraining GFM at large
$\tanb$ (the dashed lines). It is also apparent that the regions
excluded by the constraint are rather different in the
$\dll$--$\dlr$ and the $\drl$--$\drr$ planes (the
top--left and bottom--right panels respectively)
compared with the remaining four combinations. This is because,
in both cases, the contributions to the neutral Higgs penguin
interfere directly with one another, leading to the rectangular
shaped regions in the two panels. It should also be noted that
these regions are orientated at roughly ninety degrees to the $\bsg$ 
constraint in both panels. As such, the combination of the
two constraints tends to be fairly complementary when constraining
these two combinations. For the remaining four contributions, one
insertion affects the right--handed Higgs vertex while the other affects
the left--handed vertex. These contributions translate to contributions
to the unprimed and primed Wilson coefficients respectively. Direct
interference is therefore ruled out, resulting in the circular
regions found in the remaining four plots.

Let us now consider the constraint imposed by the $\bbb$ mixing
system. Here we see that, in all six panels, the constraint 
imposed by this process is comparatively mild (with the
exception of the $\drl$--$\drr$ plane). This is mainly
due to the way we impose the constraint. Since we simply
require that $\delmbs>14.5\ps^{-1}$ the regions excluded
in the panels involving the combination of an LL or an LR insertion,
and a RL or a RR insertion, tend to be rather slight due to
the large effects that are possible for these combinations
(see the end of section~\ref{CUR:bbb}). Therefore, if one
chose to impose some sort of upper bound on $\delmbs$ as
well, the constraints on SUSY flavour violation
imposed by the $\bbb$ mixing system would be far more
stringent.

Finally, let us briefly make one more remark concerning
how the plots in~\fig{Cur:mult} might change should one choose to
vary other sources of flavour violation in addition to those
already being varied in each plot. Generally, it would
be expected that quite large deviations from the contours
shown in~\fig{Cur:mult} occur once one proceeds beyond this limit,
however, there is one exception. Certain 
regions in the top--left
panel of~\fig{Cur:mult} are relatively independent of the effects
of varying $\drl$ and $\drr$. This is due to the fact that varying
$\drl$ and $\drr$ invariably only affects the primed Wilson coefficients.
The result of varying these insertions is that they, therefore, 
tend to increase the values of BR$(\bsg)$ and BR$(\bsm)$ compared
to a calculation where they are set to zero. Varying these
insertions might therefore fill in the excluded region at
the centre of the ``doughnut'' shown in the plot. The
variation, however,
will not affect (in a large manner) the regions excluded by
the outer ring that delineates the $\bsg$ constraint nor 
the regions excluded by the $\bsm$ constraint. The regions
excluded by the lower bound on $\delmbs$, however, will
tend to vary much more as varying RL and RR insertions
together with LL and LR insertions inevitably leads to
large deviations in $\delmbs$.

\FIGURE[t!]{
  \begin{tabular}{c c}
    \includegraphics[width=0.49\textwidth]{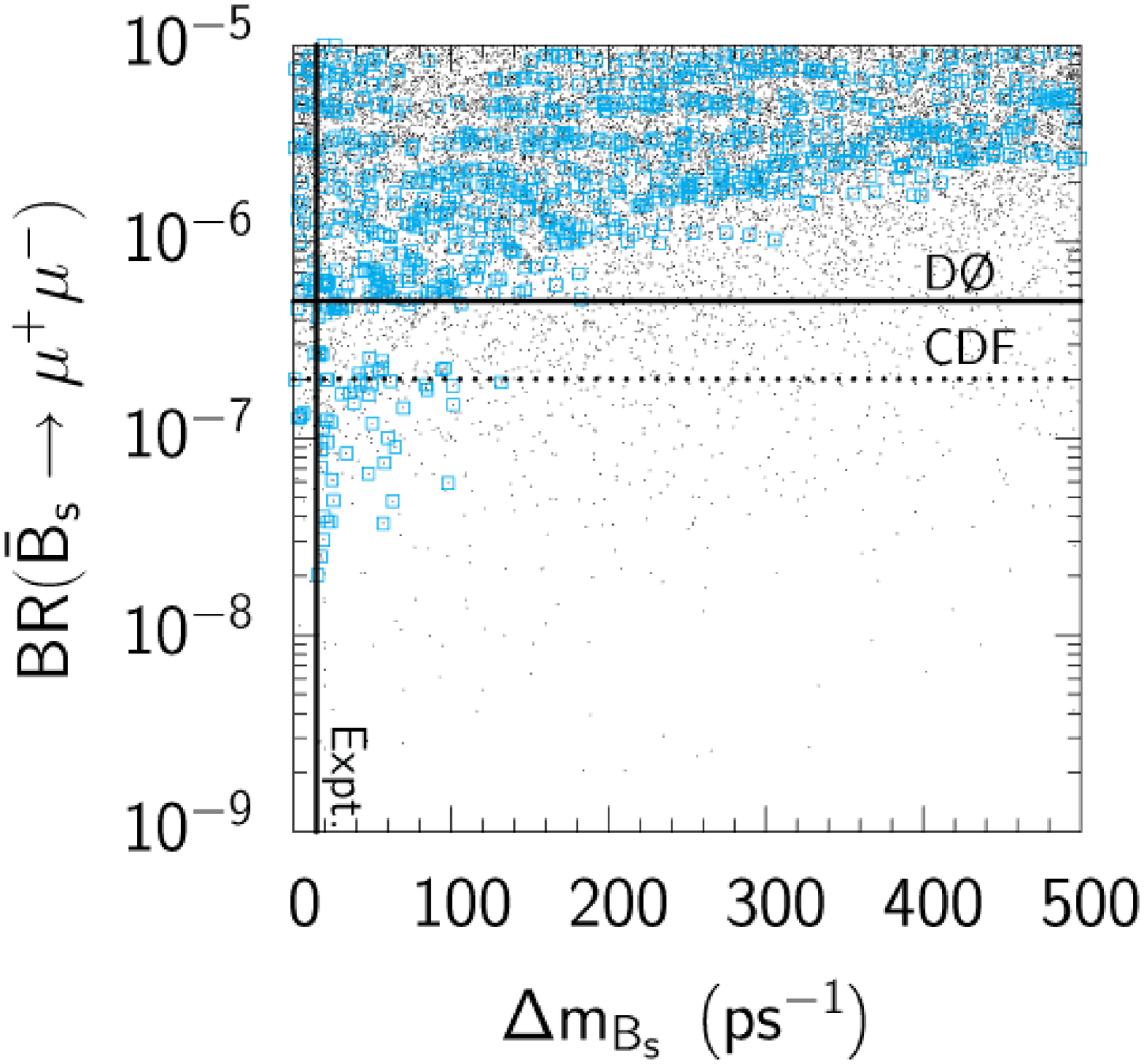}
    & \includegraphics[width=0.49\textwidth]{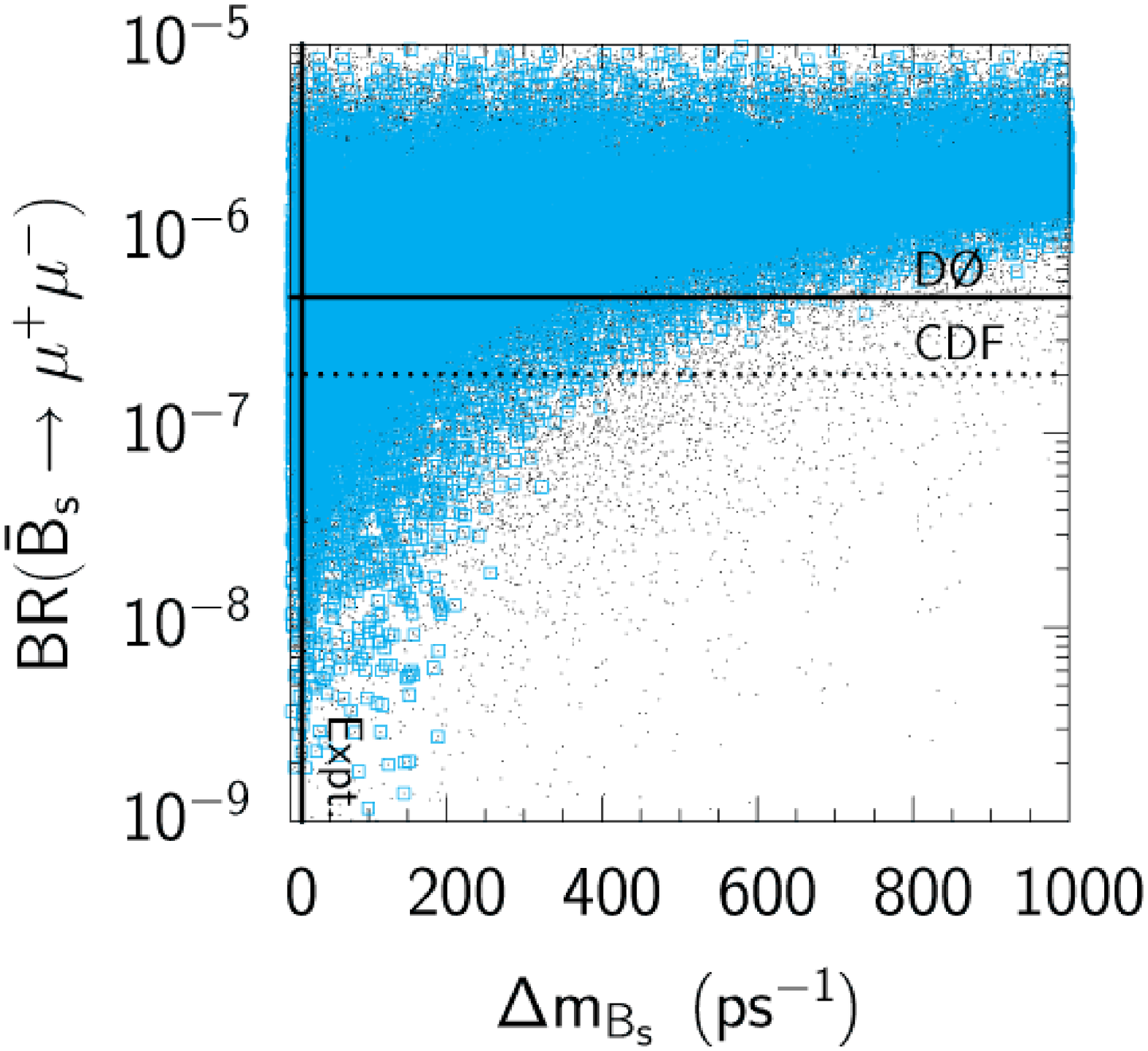}
  \end{tabular}
  \caption
  {Scatter plots illustrating the relationship between $\brbsm$
  and $\delmbs$ when all four sources of flavour violation are
  being varied. In each plot $\dll$ and $\drr$ are varied
  over the range $[-0.8,0.8]$, while the insertions $\dlr$
  and $\drl$ are varied over the range $[-0.08,0.08]$.
  The soft sector is parameterised using the relations
  $\msq=-A_u=\mgl/\sqrt{2}=\sqrt{2}\,\mu=2\,m_A$
  with $\msq=500\gev$ in the left panel and $\msq=1\tev$
  in right panel, $\tanb=40$ in both plots. Blue (grey) squares
  denote points that
  are consistent with the experimental result for $\brbsg$, while
  black dots are not.\label{Cur:Scat}}
}
The correlation between certain observables in the large
$\tanb$ regime is well known. In~\cite{BCRS:bbm},
for instance, a strong correlation between BR$(\bsm)$ and
$\delmbs$ was pointed out that exists due to the dependence
both processes have on the neutral Higgs penguin.
It was shown in~\cite{FOR:bdec}
that similar
correlations exist if one varies only one insertion at a time
in the GFM scenario.~\fig{Cur:Scat} illustrates the result
of varying all four insertions at the same time, in a particular
region of parameter space. As is evident from the figure, the
correlation between BR$(\bsm)$ and $\delmbs$ that exists
in MFV and certain limits of GFM is less pronounced once
one varies all four sources of flavour violation. This is hardly
surprising once one notices the vastly different behaviour
exhibited by the contributions arising from each insertion
to $\delmbs$~\cite{FOR:bdec}, as well as the fact
that multiple sources of flavour violation lead to exceptionally
large contributions to $\delmbs$ (see the end of section~\ref{CUR:bbb}).
The scatter plots in~\fig{Cur:Scat} also demonstrate the effect
that increasing the bounds on $\brbsm$ or $\delmbs$ might have on the
available parameter space. In the left panel (corresponding
to a relatively light SUSY spectrum and Higgs sector) it is evident that
the majority of the points consistent with $\bsg$ have already been ruled
out by the Tevatron limits on $\brbsm$. Turning to the right panel,
where the mass spectrum is twice as heavy, it is evident 
that there is far more freedom, this is of course
unsurprising as the contributions to all three processes decrease
as the mass spectrum increases. It is evident from the plot, however,
that reducing the limit on $\brbsm$ to roughly $3\times 10^{-8}$ or
observing $\delmbs$ will affect the available parameter space
dramatically.

\section{Other Constraints}
\label{CUR:other}

Before proceeding with our conclusions let us briefly discuss
how some additional constraints might affect the available
parameter space in the GFM scenario.

The first process we shall consider is
the decay $\bsll$. The effect this process might have on
the SUSY parameter space was discussed in~\cite{GHM2:bsg}
where it was shown that the combined constraints provided
by $\bsll$ and $\bsg$ indicate that the sign of the amplitude
for the decay $\bsg$ was that of the SM, unless large contributions
to the Wilson coefficients $C^{\mathrm{eff}}_9$ and
$C^{\mathrm{eff}}_{10}$ could be induced. Recalling the discussions
in section~\ref{CUR:bsg} and section~\ref{CUR:LimSing}, it is possible
for both LL and LR insertions to induce contributions to $\bsg$ that
flip the sign of the underlying amplitude (see, for example, the
upper branches of the top--left and top--right panels in~\fig{bsg:cnt}).
As we have seen, the constraints imposed by $\bsm$ and $\bbb$ mixing
typically exclude the majority of these regions at large $\tanb$.
However it is still possible, in some cases, that such regions can
remain even after one has taken into account both of these
constraints (see, for example, the top--left panel in~\fig{Cur:mult}).
In addition, it is interesting to consider how one might remove these
regions at low $\tanb$ where the $\bsm$ and $\bbb$ mixing constraints
play only a minor r{\^o}le. While a complete analysis of the $\bsll$
constraint in GFM is beyond the scope of this analysis, let us briefly
make a few comments on the possible effects of the constraint.

For LR insertions it is difficult to induce large
contributions to either $C^{\mathrm{eff}}_9$ or
$C^{\mathrm{eff}}_{10}$. This is because the operators
associated with these Wilson coefficients feature
the Dirac bilinear $s_{L}\gamma_{\mu}b_{L}$. The contributions
to these operators arising from LR insertions, therefore, appear
at second order in the mass insertion approximation and suffer
from a suppression by either the strange or bottom quark mass.
Compared with the large contributions induced by LR insertions
to the Wilson coefficient $C^{\mathrm{eff}}_{7}$ that appear
at first order in the MIA, in addition to being chirally enhanced
by the gluino mass, it is likely that the constraint imposed
by $\bsll$ will be useful in constraining the large
values of LR insertions that have effectively flipped the sign of
the $\bsg$ amplitude, at low $\tanb$ in particular. 

Turning to LL insertions,
it is possible that contributions to the operators
associated with the Wilson coefficients $C^{\mathrm{eff}}_9$ and
$C^{\mathrm{eff}}_{10}$ might be somewhat larger as contributions
proportional to $\dll$ can appear at first order in the
MIA via photon penguins~\cite{LMSS:bsll}. These contributions
are not enhanced by $\tanb$, however, and it seems likely that
for large $\tanb$ the constraint will play a similar
r{\^o}le to $\bsm$ and $\bbb$ mixing, due to the $\tanb$
enhancement of the corrections to $C^{\mathrm{eff}}_7$. The impact
the constraint might have on constraining RL and RR is less
clear as they affect the primed coefficients and the possibility
of flipping the sign of the $\bsg$ amplitude is removed. 

In summary, in most cases at large $\tanb$, the flipped sign
of the amplitude for the decay $\bsg$ is excluded by a combination
of the experimental bounds from $\bsm$ and $\bbb$ mixing. For low
values of $\tanb$ however, where the constraints provided by
$\bsm$ and $\bbb$ play only a minor r{\^o}le, it is clear that
the additional constraint provided by $\bsll$ will be
fairly useful when attempting to exclude these extreme regions
of parameter space~\cite{Silvestrini:bdec}.

Another means of constraining the GFM parameters $\dxy$ that
has been discussed in the literature is via the use of
precision electroweak observables, such as $m_W$ and
$\sin^2\theta_{\mathrm{eff}}$~\cite{HHMP:pew}. The constraints
in this case typically affect either LL or RR insertions
as contributions due to LR and RL insertions typically appear
at fourth order in the MIA (as opposed to second order for LL
and RR insertions). Taking into account these corrections,
it has been shown that, for large values of $\dll$, quite
sizeable corrections to both $m_W$ and $\sin^2\theta_{\mathrm{eff}}$
can be induced that provide a useful limit in constraining
extreme values of $\dll$~\cite{HHMP:pew}. The improved
measurements that will be made at the LHC and the next
linear collider will serve to strengthen these constraints even
more.

\section{Conclusions}
\label{CUR:conc}

In conclusion we have discussed the current limits that can be placed
on SUSY flavour violation. In particular, we have included all the
relevant $\tanb$ enhanced corrections that appear at BLO when
calculating 
the SUSY contributions to the various processes under consideration.
For the decay $\bsg$ we have reiterated the need to include BLO
corrections when deriving limits in the large $\tanb$ limit,
showing the effect of varying parameters such as $\mu$ and $\mgl$
on the BLO focusing effect discussed previously
in~\cite{OR1:bsg,OR2:bsg,FOR:bdec}. For the processes $\bsm$ and $\bbb$
mixing we have shown that useful limits can be placed on all four insertions
with the current experimental bounds. We have also combined the limits from
all three processes and investigated the effect the constraints
have in a variety of scenarios. In particular, we have shown that both
$\bsm$ and $\bbb$ mixing can play an important r{\^o}le in constraining
RL and RR insertions for $\tanb$ as low as 30. Our combined analysis
also reveals that the case $\mu<0$ can still be allowed for
small amounts of SUSY flavour violation, even for
light sparticle masses. The constraints
also serve as a useful tool when constraining multiple sources
of flavour violation and useful bounds are already available
in the $\dll-\dlr$ plane. Finally, we have discussed the prospects
available for future experiments that aim to probe the processes
$\bsm$ and $\bbb$ mixing. In particular, we have identified a pattern
of possible observations at large $\tanb$ that could only be associated
with flavour violation in the RL or RR sectors.

\acknowledgments

We would like to thank Antonio Masiero for helpful comments. J.F.
would also like to thank the HEP/PA group at Sheffield for use of
the HEPgrid cluster on which most of the numerical results of this
paper were prepared. J.F. has been supported by a PPARC
Ph.D. studentship  and the research fellowship
MIUR PRIN 2004 -- ``Physics Beyond the Standard Model''. K.O. has been
supported by the grant-in-aid for 
scientific research on priority areas (No. 441):  ``Progress in
elementary particle physics of the 21st century through discoveries
of Higgs boson and supersymmetry'' (No. 16081209) from the Ministry
of Education, Culture, Sports, Science and Techology of Japan
and the Korean government grant KRF PBRG 2002-070-C00022.


\end{document}